\documentstyle[preprint,prb,aps]{revtex} 

\begin{document}
\draft

\author{Robert H. Gowdy\thanks{%
E-mail address: (Internet) rgowdy@cabell.vcu.edu}  \\ 
Department of Physics,\\ Virginia Commonwealth University,\\ Richmond, VA
23284-2000}

\title{Affine Projection Tensor Geometry: Decomposing the Curvature Tensor
When the Connection is Arbitrary and the Projection is Tilted}

\date{November 19, 1993}

\maketitle

\begin{abstract}
This paper constructs the geometrically natural objects which are associated
with any projection tensor field on a manifold with any affine connection.
The approaches to projection tensor fields which have been used in general
relativity and related theories assume normal projection tensors of
co-dimension one and connections which are metric compatible and
torsion-free. These assumptions fail for projections onto lightlike curves
or surfaces and other situations where degenerate metrics occur as well as
projections onto two-surfaces and projections onto spacetime in the higher
dimensional manifolds of unified field theories. This paper removes these
restrictive assumptions. One key idea is to define two different ''extrinsic
curvature tensors'' which become equal for normal projections. In addition,
a new family of geometrical tensors is introduced: the cross-projected
curvature tensors. In terms of these objects, projection decompositions of
covariant derivatives, the full Riemann curvature tensor and the Bianchi
identities are obtained and applied to perfect fluids, timelike curve
congruences, string congruences, and the familiar 3+1 analysis of the
spacelike initial value problem of general relativity.

PACS:
\end{abstract}
\pacs{04.20.Cv, 02.40.Hw, 04.30.+x, 04.50.+h}

\section{Introduction}

Applications of general relativity and related theories can often be stated
in terms of projection tensor fields. Familiar examples are the co-moving
frame projections which are central to hydrodynamics\cite
{tcong.prjctn,co-moving,intrinsic.geom.a,intrinsic.kv.geom,intrinsic.geom,%
intrinsic.geom.I,intrinsic.geom.II,intrinsic.geom.III,tcong.q.geom}
and the 3-surface projections which arise in initial value problems.\cite
{iv-projectns} In situations where projection tensors are not traditionally
used, they often provide an improved description. For example, the embedding
in a metric space of a submanifold with co-dimension higher than one is
usually described by a non-unique set of normal vectors\cite{subman-norms}
but is better described by the unique normal projection into the subspace
tangent to the submanifold.

Aside from elegance and improved invariance properties, projection tensor
techniques offer another advantage: They always lead to the same operations
and those operations are always simplified by the construction of the same
geometrical objects regardless of the nature of the system which is being
described. Thus, I am led to describe a {\em projection tensor geometry}
which contains results of wide applicability. The resulting geometry greatly
enlarges the scope of projection tensor methods because it does not assume
projections onto surfaces or normal ({\it i.e.} perpendicular) projections,
or even the existence of a metric tensor.

When only one projection-tensor field is considered, this geometry is
modeled on traditional surface embedding theory and generalizes the
intrinsic and extrinsic curvature of a surface to the case of a
projection-tensor field which need not be surface-forming. Just as in
surface embedding theory, the main result is a decomposition of the
Riemannian curvature tensor in terms of the projection curvatures. When a
projection-tensor field is hyper-surface forming, the curvature
decomposition includes the Gauss-Codazzi equations\cite{Gauss-Codazzi} which
have become familiar to relativists as the foundation of the 3+1
decomposition of the spacelike hypersurface initial value problem in general
relativity. In familiar cases where the projection-tensor field is not
surface-forming --- fluid flow for example --- the projection curvatures
turn out to be composed of such well-known quantities as the shear,
divergence, and vorticity of fluid flow and the curvature decomposition
leads to such familiar results as the Raychaudhuri equation.\cite
{Raychaudhuri.a,Raychaudhuri} In less familiar cases such as two-dimensional
projections in a four-dimensional manifold, projections in the higher
dimensional manifolds of unified field theories, and projections onto
lightlike curves and surfaces, the curvature decomposition introduces
geometrical objects and relationships which I, at least, have not seen
before.

A previous paper on this subject introduced a compact 'decorated index'
formalism for describing a single projection tensor field and applied it to
the hydrodynamics and thermodynamics of a perfect fluid in general relativity.%
\cite{prjctn1} That paper imposed two major restrictions on the situations
which it could cover: (1) There had to be a metric tensor which was at least
invertible, thus excluding a projection-tensor approach to spacetime
perturbation theory which was developed in my earlier papers\cite
{pert1,pert2} as well as any discussion of unified field theories which use
non-metric-compatible connections. (2) Projections in null or light-like
directions were not allowed, thus excluding a projection-tensor approach to
the propagation of radiation and the characteristic initial value problem.
This paper removes those restrictions.

Although most of the applications which I have in mind involve spaces with
zero torsion, I carry the torsion tensor throughout. As has often been
observed, particularly in connection with the ECSK theory of gravity,
differential geometry is a far more elegant and symmetrical theory with the
torsion tensor present than without it.\cite{ECSKrefs} Here, I find it
useful to define generalized torsion tensors associated with a projection
tensor field in order to produce projected structure equations which are
simple and symmetrical.

The compact index notation of the previous paper is not easily generalized
to multiple projection tensor fields. Since multiple projection tensors
often arise in applications, this paper will mostly use the familiar,
unadorned index notation of tensor analysis. Although clumsy in some ways,
this notation is one which we can all understand without explanations which
might obscure the essential points which I am trying to make. I depart
slightly from the notation in my previous papers by using an operator
notation for covariant derivatives: $\nabla _\delta {}M^{\alpha \beta
}\!_{\mu \nu }{}=M^{\alpha \beta }\!_{\mu \nu ;\delta }$. Notice that this
operator notation does not change the convention that the differentiating
index is added to the {\em end} of the list of tensor indexes. Also notice
the distinction between a covariant derivative, which increases the rank of
a tensor, and a directional derivative (used in my previous papers) which
does not. The spacetime signature is taken to be $-+++$ and my conventions
on the torsion and curvature tensors may be seen in Eqs. (\ref
{torsion.def},\ref{curv.def}) .\cite{curv.note}

Section \ref{notation} of this paper reviews the basic properties of
projection tensor fields and defines the new geometrical structures and
operations which become natural when a projection tensor field is present.
Section \ref{settings} introduces a few of the many situations in which
projection tensor fields play central roles. The key formal results of the
paper are contained in Section \ref{prjctn.curv} which defines the
generalized projection curvatures and in Section \ref{gstructure} which
presents the projection decompositions of the metricity and torsion tensors
as well as the Riemann and Ricci tensors . This section introduces several
new tensor fields --- cross-projected torsion and curvature tensors ---
which are needed for a full analysis of the way that a projection tensor
field interacts with a connection. A corresponding set of projected Bianchi
identities obeyed by these tensors is also worked out. Section \ref
{applications} shows how these results are used in two familiar situations,
fluid dynamics and the spacelike initial value problem of general
relativity. The applications considered here are taken just far enough to
demonstrate and provide a familiar context for the techniques developed in
this paper. I expect to return in later papers to the new applications which
these techniques make possible.

\section{Projection Tensor Definitions}
\label{notation}

\subsection{Projection Tensor Fields}

A projection tensor-field $H$ assigns to each point $P$ of a manifold a
linear map of the tangent space $H\left( P\right) :T_P\rightarrow T_P$ such
that
\begin{equation}
H^2=H.\label{prjctn.def}
\end{equation}
It follows from this definition and the basic properties of a vector space
that the projection tensor $H$ acts as an identity operator on the
projection subspace $HT_P$.

If $H$ is a projection tensor field, and $I$ is the identity map, the tensor
field
\begin{equation}
V=I-H \label{comp.def}
\end{equation}
is also a projection tensor field which will be called the {\em complement}
of $H$. An immediate consequence of Eqs. (\ref{prjctn.def},\ref
{comp.def}) is $VH=HV=0$. The complement of an expression involving
projection tensor fields is obtained by replacing each projection tensor by
its complement. As was discussed in an earlier paper\cite{prjctn1},
complementation is a valuable feature because the complement of a definition
or identity is always another valid definition or identity, thus cutting in
half the work of stating or deriving expressions.

The natural action of a projection tensor on the cotangent spaces to a
manifold is defined by its pull-back $H^{*}$. In terms of components, if $%
\beta $ is a one-form with components $\beta _\alpha $ and $u$ is a vector
with components $u^\alpha $
$$
\left( Hu\right) ^\alpha =H^\alpha \!_\rho {}u^\rho ,\qquad \left(
H^{*}\beta \right) _\alpha =H^\rho \!_\alpha {}\beta _\rho .
$$
Since higher rank tensors may always be regarded as linear functions of
forms and vectors, projection tensors can act on them by acting on their
arguments. For example, the tensor with components $M^{\alpha \beta }\!_{\mu
\gamma }{}$ would have a number (sixteen to be precise) of projections by
the tensor $H$ and its complement, including the projection
\begin{equation}
\begin{array}{c}
M\left[ ^{V~H}{}_{H~V}\right] {}^{\alpha \beta }\!_{\mu \gamma }=
V^{\,\alpha }\!_\sigma {}H^{\,\beta }\!_\rho {}\,H^{\,\tau }\!_\mu
{}\,M^{\sigma \rho }\!_{\tau \nu }{}\,V^{\,\nu }\!_\gamma .
\end{array}
\label{tensor.prjctn.notn}
\end{equation}
Notice the way in which I have named this tensor projection. I will often
use this sort of naming convention for projected tensors and tensor
subspaces so that one can see at a glance which projections have been
performed on which index positions.

\subsection{Projection Subspaces and Projection Identities}

A projection tensor field such as $H$ splits each tensor space into {\em %
projection subspaces}. The tangent space $T_P$ at the point $P$ is split
into the projection subspaces $HT_P$ and $VT_P$. The cotangent space $\hat
T_P$ is split into the projection subspaces $H^{*}\hat T_P$ and $V^{*}\hat
T_P$. Similarly a tensor space such as $T_P\otimes T_P\otimes \hat
T_P\otimes \hat T_P$ is split into sixteen projection subspaces, including
the one inhabited by the example discussed above:
\begin{equation}
T\,\left[ ^{V\;H}{}_{H~V}\right] _P=VT_P\otimes
HT_P\otimes H^{*}\hat T_P\otimes H^{*}\hat T_P.
\label{tensor.prjctn.subsp}
\end{equation}
An object which lies entirely in a projection subspace at each point of a
manifold is said to be {\em fully projected} and is said to obey a {\em %
projection identity}. For example, a vector $v$ which is in the projection
subspace $HT_P$ obeys the identity $Hv=v$. Similarly, a fully projected
tensor $M_{HV}^{VH}$ which is in the projection subspace $T\,\left[
^{V\;H}{}_{H~V}\right] _P$ defined in Eq. (\ref{tensor.prjctn.subsp})
obeys the projection identity%
$$
M_{HV}^{VH}\left[ ^{V\;H}{}_{H~V}\right] {}^{\alpha \beta }\!_{\mu \gamma
}{}=V^{\,\alpha }\!_\sigma \,{}H^{\,\beta }\!_\rho \,{}H^{\,\tau }\!_\mu
\,{}M_{HV}^{VH\,}{}^{\sigma \rho }\!_{\tau \nu }\,{}V_{\;\;\gamma }^\nu
{}={}M_{HV}^{VH\,}{}^{\alpha \beta }\!_{\mu \gamma }.
$$

Notice the distinction between a tensor $M_{HV}^{VH}$ which inhabits a
particular projection subspace and a tensor $M\left[ ^{V~H}{}_{H~V}\right] $
which is the result of projecting a tensor $M$ into that subspace. The
subscripts and superscripts which identify a fully projected tensor such as $%
M_{HV}^{VH}$ will be called {\em projection labels}. Often these labels will
be replaced by {\em variable projection labels} which stand for possible
choices of projections. For example, $M_{ZW}^{XY}$ with the choices $\left(
X,Y,Z,W\right) =\left( H,V,V,H\right) $ would stand for the tensor $%
M_{VH}^{HV}$. In order to avoid confusion between projection labels and
tensor indexes, the labels are grouped together in a block immediately after
the symbol for the tensor and the first tensor index is positioned to the
right of the last projection label as in $R_{HH}^V{}_\alpha {}^\beta {}_{\mu
\nu }{}$.

\subsection{Adapted Frames}

For each projection tensor field $H$ one can define an {\em adapted
reference frame} which assigns the vectors $e_a{}\left( P\right)
,e_A{}\left( P\right) $ to each point $P$ with the vectors $\left\{
e_a{}\right\} $ spanning the subspace $HT_P$ and the vectors $\left\{
e_A{}\right\} $ spanning the subspace $VT_P$. Because $H$ acts as an
identity operation on $HT_P$ and $V$ similarly acts as an identity on $VT_P$%
, the two sets of basis vectors can be characterized by
$$
H{}e_a{}=e_a{},\qquad Ve_A{}=e_A{}.
$$
and the non-zero adapted-frame projection tensor components are%
$$
H^a\!_b{}=\delta _b^a,\qquad V^A\!_B{}=\delta _B^A{}.
$$
In an adapted frame, the fully projected tensor $M_{HV}^{VH}{}^{\alpha \beta
}{}_{\mu \nu }$ which I have been using as an example would have non-zero
components $M_{HV}^{VH\,}{}^{Ab}{}_{cD}$ with all other components, such as $%
M_{HV}^{VH}\!\,^{ab}\!_{cd}$ equal to zero.

\section{Settings of Projection Tensor Geometry}
\label{settings}

\subsection{Normal Projection Tensors}

So long as there is a regular metric tensor, one can define a {\em normal
projection tensor field} to be a projection $H$ such that the kernel of $H$
is orthogonal to $Hv$ for any vector $v$. In other words, one requires $%
g\left( Hv,u\right) =0$ whenever $Hu=0$ or%
$$
H_{\beta \alpha }{}v^\alpha {}u^\beta =0\text{ whenever }H^\alpha \!_\beta
{}u^\beta {}=0.
$$
It is easy to see that a sufficient condition for $H$ to be normal is $%
H_{\alpha \beta }{}=H_{\beta \alpha }{}$. To show that the condition is also
necessary, notice that a reference frame which is adapted to a normal
projection tensor satisfies g$\left( e_a,e_A\right) =g_{aA}=0$ so that the
non-zero adapted frame components of $H_{\alpha \beta }$ are $%
H_{ab}=g_{ar}{}H^r\!_b=g_{ar}{}\delta _b^r=g_{ab}$ and are manifestly
symmetric.

A remark is needed at this point: Many discussions of projection tensor
applications begin with the relation $H_{ab}=g_{ab}$. This relation is
correct {\em only} for normal projection tensors. As will be seen in the
next section, it cannot be imposed on null projection tensors or the
projection tensors associated with spacetime deformations.

Normal projection tensors are useful because they are uniquely determined
from the projection subspace $HT_P$ of the tangent space $T_P$ at each point
of a manifold. For example, given a spacelike hypersurface, there are two
(past and future-pointing) unit normal vectors $n^\alpha $ at each point.
For either one, the projection which takes arbitrary vectors into vectors
tangent to the surface is just
\begin{equation}
H^\alpha \!_\beta =\delta _\beta ^\alpha +n^\alpha
n_\beta .
\label{hypsurf.prjctn}
\end{equation}
For a more general submanifold, one would choose an orthonormal basis on the
subspace of vectors normal to the surface at each point and use this set of
normal vectors to construct a unique projection.

A slightly different example shows that normal projection tensors need not
be projections onto the tangent spaces of submanifolds. A fluid can be
described by giving the four-velocity $u^\alpha $ at each place and time.
The constraint $u_\alpha u^\alpha =-1$ then ensures that the tensor%
$$
V^{\,\alpha }\!_\beta =-u^\alpha u_\beta
$$
obeys the requirement $V^2=V$ and is a projection tensor. In this case, $V$
projects onto the tangent spaces to the fluid world-lines. However the story
is different for the complementary projection tensor $H=I-V$ or%
$$
H^\alpha \!_\beta =\delta _\beta ^\alpha +u^\alpha u_\beta
$$
which takes the space components of vectors in the local rest frame of the
fluid. When the fluid has twist or vorticity, these local rest frames cannot
be integrated to give a global rest-frame. Thus, the normal projection
tensor $H$ may not be surface-forming.

One idea which will be revisited in section\ref{gstructure} is Barry
Collins' notion of the intrinsic geometry which is associated with a
non-surface-forming projection tensor field.\cite{intrinsic.geom} A similar
idea, the quotient geometry which results from a single Killing vector
field, was developed by Geroch in a framework very similar to the one used
here.\cite{intrinsic.kv.geom}

For these normal projection tensors, the techniques which are developed in
this paper reduce to familiar calculations. However, there are some new
insights to be gained from seeing these old calculations in this more
general setting, so this paper reviews them in section \ref{applications}.

\subsection{Null Projection Tensors}
\label{null}

In general, a {\em null projection tensor} $H$ is characterized by a
projected metric tensor $g\left[ _{HH}\right] $ (or projected inverse metric
tensor $g^{-1}\left[ ^{H\;H}\right] $) which is non-invertible and thus
fails to define a metric on the subspace $HT_P$. In spacetime, a null
projection tensor is one which preserves exactly one null ray.

For a projection onto a {\em null hypersurface}, $H$ could take the form
\begin{equation}
H^\alpha \!_\beta {}=-\ell ^\alpha {}n_\beta
+e_1{}^\alpha {}\,e_{1\,\beta }+e_2{}^\alpha \,e_{2\,\beta }
\label{norm.null.prjctn}
\end{equation}
where $\ell $ is a null vector tangent to the hypersurface and $n$ is a null
vector pointing out of the surface. I will not go into detail here, but it
is evident that such a projection tensor is not symmetric and cannot be
normal. Thus, the usual definitions of surface curvature do not work. One is
left to guess whether the extrinsic curvature of such a surface should be
built from derivatives of $\ell $ or from derivatives of $n$. The case for $%
n $ is that it points out of the surface like the normal to a spacelike
surface does. The case for $\ell $ is that an attempt to approximate a null
surface by a sequence of Lorentz-boosted spacelike surfaces shows their
normal rays approaching the null ray defined by $\ell $.

The next section will show that there are two, equally natural, definitions
of the curvature of a projection-tensor field. For normal projection tensor
fields, they coincide. For null projection tensor fields they do not
coincide and one turns out to be the $n$ version of the extrinsic curvature
while the other is the $\ell $ version. Thus, the general projection-tensor
geometry developed in this paper adapts easily to the peculiarities of null
projection tensors.

One situation which is naturally described by a system of null projection
tensors is radiation originating from a compact source. Far from the source,
slice spacetime by null hypersurfaces which look like future light cones.
Each of these null hypersurfaces contains a congruence of null geodesics
which correspond, in the optical limit, to world-lines of the radiation.
Within this setting one formulates the characteristic initial value problem
to describe the propagation of radiation. This null-surface formulation is
particularly useful for numerical computations because it lends itself to a
conformal transformation which permits a finite evaluation grid to reach
future null infinity where precise definitions of the amount of
gravitational radiation flux are available.\cite
{char.in.val.rad,char.in.val.rad.a,cauch.char.init.val}

The usual approaches to the characteristic initial value problem introduce
spinors, null tetrads, or pairs of null congruences and lead to equations of
motion for the corresponding connection coefficients and components of the
Weyl tensor\cite{char.in.val,char.in.val.a,2null.calculus}. A
projection-tensor approach would begin with the observation that there is
not just one projection tensor field in this system but a nested pair of
them --- the projections onto the level surfaces of the optical function and
projections onto the null geodesics which lie in those level surfaces. This
nested set of projections with their corresponding intrinsic and extrinsic
curvatures is not fully exploited in any analysis that I am aware of.

This paper extends the definitions of intrinsic and extrinsic curvatures so
that they can be applied to a null-hypersurface analysis of radiation.\cite
{ext.curv.defs.gen} That analysis will be developed in a later paper.

\subsection{Deformation Geometries}

Given a family of manifolds, each carrying a metric and a connection, and
each labeled by a set of parameters $\left\{ \epsilon ^A\right\} ,$ regard
each manifold $S\left( \epsilon \right) $ as a submanifold of a larger
manifold, $M$, and regard the parameters as functions on that larger
manifold. When the submanifolds are spacetimes, I call the larger manifold a
{\em spacetime deformation}. Geroch used this concept to provide a
geometrical framework for discussing limits of spacetime sequences.\cite
{geroch.defs} I have used it as a geometrical framework for spacetime
perturbation theory.\cite{pert1,pert2}

At each point of the larger manifold $M$ choose a basis $\left\{ e_\alpha
\right\} $ for the subspace tangent to the submanifold $S\left( \epsilon
\right) $. Let $g^{\alpha \beta }$ be the corresponding components of the
spacetime metric $S\left( \epsilon \right) $ and construct the tensor field $%
g^{-1}=g^{\alpha \beta }{}e_\alpha {}\otimes e_\beta $ on $M$. This tensor
field plays the role of an inverse metric tensor on $M$. It can be used to
map forms into vectors according to
$$
g^{-1}\left( \mu \right) =g^{\alpha \beta }{}e_\alpha {}\otimes e_\beta
{}\left( \mu \right) =e_\alpha {}g^{\alpha \beta }{}\mu _\beta =e_\alpha
{}\mu ^\alpha
$$
and of course it provides an inner product for forms according to%
$$
\mu \cdot \nu =g^{-1}\left( \mu \right) \cdot \nu =g^{\alpha \beta }{}\mu
_\alpha {}\nu _\beta .
$$
However, when this tensor is applied to the differential forms $d\epsilon ^A$%
, it maps them to zero. Thus it provides a {\em degenerate inner product}
and cannot be inverted to give a metric on the larger manifold.

The geometry of a spacetime deformation is incomplete unless one adds some
structure to it. The approach which might at first seem obvious, completing
the metric tensor by adding terms corresponding to the parameter directions
is not very useful. It adds arbitrary parameter-space structure which has
nothing to do with understanding the spacetimes in the deformation. Geroch's
approach is to add a vector field which maps the points of a spacetime to
the points of its neighbors.\cite{geroch.defs} My own work imposes a
projection tensor field $H$ which basically does the same job as Geroch's
vector field. I require the projected tangent space $HT_P$ at each point of
the deformation to be the tangent space to the spacetime which passes
through $P$.\cite{pert1} Since there is no regular metric on the
deformation, this requirement leaves some freedom to choose the projection
tensor field -- a freedom which corresponds to Geroch's choice of vector
field and to the choice of gauge in perturbation theory.

A major point of my approach to deformation theory is that it can be modeled
on a projection-tensor formulation of surface embedding theory. However,
without a regular metric on the deformation, a direct link with the
Gauss-Weingarten theory of surface embedding is lacking. This paper provides
that direct link.

\subsection{Geometrical Symmetry Breaking in Unified Field Theories}

Many promising approaches to unified field theory involve manifolds with
more than four dimensions which evolve in such a way that all but four of
the dimensions ''collapse'' or else fail to expand.\cite{K.K.symmetry.break}
Regardless of the details of the theory or the precise symmetry-breaking
mechanism, the final state is best described by a {\em spacetime
projection-tensor field} which projects to zero those vectors which point in
''collapsed'' directions. When this description is used, all of the
dynamical fields which appear in the resulting four-dimensional spacetime
become components of the geometrical objects which are discussed in this
paper.

All of the identities and field equations which can arise in this type of
unified field theory are implicit in the projection tensor geometry
developed by this paper. Further, they are worked out in a unified
geometrical framework which should provide new insights into these theories.
I expect to explore some of these insights in later papers.

\section{Generalized Projection Curvatures}
\label{prjctn.curv}

\subsection{Projection Curvatures Without a Metric}

Given a projection tensor $H$ and a connection, the tensor
\begin{equation}
h_H{}^\alpha {}_{\gamma \delta }{}=H^\rho \!_\gamma
{}H^\sigma \!_\delta {}\nabla _\sigma {}H^\alpha \!_\rho
\label{prjctn.curv.def}
\end{equation}
is defined to be the {\em curvature} of $H$. In the familiar case of a
spacelike hypersurface with a well-defined normal vector $n^\alpha $,
Eq. (\ref{hypsurf.prjctn}) can be substituted into this definition to
relate it to the familiar expression for the {\em extrinsic curvature} or
{\em second fundamental form} $k_{\gamma \delta }$ of the surface.\cite
{ext.curv.defs,ext.curv.defs.a,ext.curv.defs.b,ext.curv.defs.gen}%
$$
h_H{}^\alpha {}_{\gamma \delta }{}=H^\rho \!_\gamma {}H^\sigma \!_\delta
{}n^\alpha \,{}\nabla _\sigma {}n_\rho {}=n^\alpha \,{}k_{\gamma \delta }{}.
$$

The definition of the projection curvature tensor is projected explicitly on
two of its three indexes. However, it is not difficult to show that it obeys
the full set of projection identities
\begin{equation}
h_H\left[ ^V{}_{H~H}{}\right] {}^\alpha {}_{\gamma
\delta }{}=V^\alpha \!_\tau {}h_H{}^\tau {}_{\rho \sigma }{}H^\rho \!_\gamma
{}H^\sigma \!_\delta {}=h_H{}^\alpha {}_{\gamma \delta }{}.
\label{h.prjctn.ids}
\end{equation}
This result is obtained by taking the covariant derivative of Eq. (\ref
{prjctn.def}), the defining requirement for $H$ to be a projection tensor,
and then projecting the result.

Another way to project the covariant derivative of a projection tensor
yields the tensor
\begin{equation}
\begin{array}{c}
h_H^T{}_\gamma {}^\alpha \!_\delta {}=
H^\alpha \!_\rho {}{}H^\sigma \!_\delta {}{}\nabla _\sigma {}{}H^\rho
\!_\gamma {}.
\end{array}
\label{prjctn.transp.curv.def}
\end{equation}
When a metric is available for raising and lowering indexes, this tensor is
the curvature associated with the transpose of $H$. Because of this
association with the transposed projection, this tensor will be called the
{\em transpose curvature} of $H$. For a normal projection tensor, it is
exactly the same as the tensor $h_H$ with the indexes appropriately raised
and lowered. By differentiating and projecting Eq. (\ref{prjctn.def}),
it is a straightforward matter to show that $h_H^T$ obeys the projection
identity
\begin{equation}
h_H^T\left[ _V{}^H{}_{H\,}\right] {}_\gamma {}^\alpha
\!_\delta {}=h_H^T{}_\gamma {}^\alpha \!_\delta {}
\label{hT.prjctn.ids}
\end{equation}

In addition to the two curvature tensors $h_H$ and $h_H^T$ associated with
the projection tensor $H$, the same definitions yield a projection curvature
$h_V$ and transpose curvature $h_V^T$ associated with the complementary
tensor $V$. These curvature tensors obey projection identities which are
simply the complements of Eqs. (\ref{h.prjctn.ids},\ref{hT.prjctn.ids}%
):
\begin{equation}
h_V\left[ ^H{}_{V\,~V}{}\right] {}^\alpha {}_{\gamma
\delta }{}=h_V{}^\alpha {}_{\gamma \delta }{},\qquad h_V^T\left[
_H{}^V{}_V\right] {}_\gamma {}^\alpha \!_\delta {}=h_V^T{}_\gamma {}^\alpha
\!_\delta {}.
\label{hcomp.prjctn.ids}
\end{equation}
In the familiar case where $H$ projects onto a family of spacelike
hypersurfaces with a unit timelike normal vector field $n$, these two
curvatures are the same and are easily found by using $V^{\,\alpha }\!_\beta
{}=-n^\alpha {}n_\beta $ in place of $H^\alpha \!_\beta $ in Eq. (\ref
{prjctn.curv.def}).%
$$
h_V^T{}^\alpha {}_{\gamma \delta }{}=h_V{}^\alpha {}_{\gamma \delta
}{}=V^{\,\rho }\!_\gamma {}V^{\,\sigma }\!_\delta {}{}\nabla _\sigma
{}V^{\,\alpha }\!_\rho {}{}=-a^\alpha {}V_{\gamma \delta }{}
$$
where $a^\alpha {}=n^\sigma \,{}\nabla _\sigma {}{}n^\alpha {}$ is the
acceleration (or curvature) of the hypersurface-orthogonal world lines.

\subsection{Decomposition of the Projection Gradient}

The covariant derivative $\nabla H$ of a projection tensor $H$ will arise
whenever one takes the covariant derivative of a tensor which obeys
projection identities. Since I am engaged in expressing everything in terms
of tensors which obey projection identities, the projection gradient $\nabla
H$ will arise often. Thus, my first task is to express the projection
gradient in terms of fully projected tensors. The resulting expression will
be the key to everything else in this paper, so I will include more of the
details of its derivation than I would for a result of lesser significance.

Use the decomposition of the identity tensor $I=H+V$ to force a
decomposition of $\nabla H$ which I write symbolically as%
$$
\nabla H=\nabla H\left[ ^I\,_{I~I}\right] =\nabla H\left[
^{H+V}\,_{H+V~~H+V}\right]
$$
In this same abbreviated notation, the various projection curvatures (with $%
\delta $ the differentiating index) are%
$$
h_H{}^\alpha {}_{\gamma \delta }{}=\nabla H\left[ ^I\,_{H~H}\right]
{}^\alpha \!_{\gamma \delta }{},\qquad h_H^T{}_\gamma {}^\alpha \!_\delta
{}=\nabla H\left[ ^H\,_{I~H}\right] {}^\alpha \!_{\gamma \delta }{}
$$
$$
h_V{}^\alpha {}_{\gamma \delta }{}=-\nabla H\left[ ^I\,_{V~V}\right]
{}^\alpha \!_{\gamma \delta }{},\qquad h_V^T{}_\gamma {}^\alpha \!_\delta
{}=-\nabla H\left[ ^V\,_{I~V}\right] {}^\alpha \!_{\gamma \delta }{}
$$
and the decomposition becomes%
$$
\begin{array}{c}
\nabla _\delta {}H^\alpha \!_\gamma {}=h_H\left[ ^H\,_{I~I}\right] {}^\alpha
{}_{\gamma \delta }{}+\nabla H\left[ ^H\,_{H~V}\right] {}^\alpha \!_{\gamma
\delta }{}+h_H^{T\;}\left[ _V{}^I\,_I\right] {}_\gamma {}^\alpha \!_\delta
{}-h_V\left[ ^H{}_{I~I}\right] {}^\alpha {}_{\gamma \delta }{} \\
+h_H\left[ ^V{}_{I~I}\right] {}^\alpha {}_{\gamma \delta }{}-h_V^{T\;}\left[
_H{}^I\,_I\right] {}_\gamma {}^\alpha \!_\delta {}+\nabla H\left[
^V\,_{V~H}\right] {}^\alpha \!_{\gamma \delta }{}-h_V^{T\;}\left[
_V{}^I\,_I\right] {}_\gamma {}^\alpha \!_\delta {}
\end{array}
$$
The two terms which have not been expressed in terms of curvatures can be
shown to vanish by projecting the covariant derivative of Eq. (\ref
{prjctn.def}). Half of the remaining terms vanish because of the projection
identities given in Eqs. (\ref{h.prjctn.ids},\ref{hT.prjctn.ids},\ref
{hcomp.prjctn.ids}). The remaining projections have no effect because of the
same projection identities. The resulting decomposition is just
\begin{equation}
\nabla _\delta H^\alpha \!_\gamma {}=h_H{}^\alpha
{}_{\gamma \delta }{}-h_V{}^\alpha {}_{\gamma \delta }{}+h_H^T{}_\gamma
{}^\alpha \!_\delta {}-h_V^T{}_\gamma {}^\alpha \!_\delta {}.
\label{prjctn.grad.decomp}
\end{equation}
The decomposition of the complementary projection gradient $\nabla $$V$ is
given by the complement of this expression --- Exchange $H$ and $V$
everywhere.

Because each projection curvature tensor has two indexes which project into
the same subspace, one can either contract those two indexes or else extract
symmetric and antisymmetric parts. In the absence of a metric, one can
define the {\em divergence form }%
$$
\theta _H^T{}_\gamma {}=h_H^T{}_\gamma {}^\alpha \!_\alpha {},
$$
the {\em twist or vorticity\ tensor}
$$
\omega _H{}^\alpha {}_{\mu \nu }\!=\frac 12\left( h_H{}^\alpha {}_{\mu \nu
}{}-h_H{}^\alpha {}_{\nu \mu }{}\right)
$$
and the {\em expansion rate tensor}
$$
\theta _H{}^\alpha {}_{\mu \nu }\!=\frac 12\left( h_H{}^\alpha {}_{\mu \nu
}{}+h_H\!^\alpha {}_{\nu \mu }{}\right) .
$$
For fluid flow characterized by a tangent vector field $u^\alpha {}$, and $%
H=\delta _\beta ^\alpha {}+u^\alpha {}u_\beta {}$, these definitions are
closely related to the usual definitions of vorticity, divergence, and
expansion rate:%
$$
\omega _H{}^\alpha {}_{\mu \nu }{}=u^\alpha {}\omega _{\mu \nu }{},\quad
\theta _H{}^\alpha {}_{\mu \nu }{}=u^\alpha {}\theta _{\mu \nu }{},\quad
\theta _H^T{}_\gamma {}=u_\gamma {}\theta .
$$

When a metric tensor is present, it can be used to raise or lower indexes
and define the additional quantities: $\theta _H\,^\alpha {}$, $\omega
_{H\,}^T{}^{\alpha \,\mu \nu }{}$, $\theta _{H\,}^T{}^{\alpha \,\mu \nu }{}$%
, as well as the shear tensors%
$$
\sigma _H{}^\alpha {}_{\mu \nu }{}=\theta _H{}^\alpha {}_{\mu \nu }{}-\frac
1{d_H}\theta _H\,^\alpha {}H_{\mu \nu }{},\qquad \sigma _H^T{}^{\alpha \,\mu
\nu }{}=\theta _H^T{}^{\alpha \,\mu \nu }{}-\frac 1{d_H}\theta _H^T\,^\alpha
{}H^{\mu \nu }{}
$$
where $d_H=H^\rho \!_\rho {}$ is the dimensionality of the projected
subspace $HT_P$.

\subsection{The Projected Connection}
\label{prjctd.cnctn}

\subsubsection{Projected and Anti-projected
Covariant Derivatives}
\label{prjctd.antiprjctd}

Consider a vector field $v$ such that $v\left( P\right) \in HT_P$ for every
point $P$ in a manifold. This vector field obeys the identity $Hv=v$. The
covariant derivative $\nabla v$ of this field can be thought of as having a
part which primarily reflects the behavior of the projection tensor and a
part which reflects the behavior of $v$ within the projected subspaces. The
part of the covariant derivative which reflects how $v$ changes within the
projected subspaces is the {\em projected covariant derivative} $Dv$ with
components
\begin{equation}
D_\delta {}v^\alpha {}=H^\alpha \!_\rho {}\nabla
_\delta {}v^\rho \!.
\label{prjctd.deriv.1}
\end{equation}
The part of the covariant derivative which ignores how $v$ changes within
the projected subspace is the {\em anti-projected covariant derivative} $%
\bar D{}v$ with components
\begin{equation}
\bar D_\delta {}v^\alpha {}=V^\alpha \!_\rho
{}\nabla _\delta {}v^\rho \!
\label{anti.prjctd.deriv.1}
\end{equation}
More generally, if $M$ is a fully projected tensor obeying the projection
identities $OM=M,$ the projected covariant derivative takes the form $%
DM=O\left( \nabla M\right) $ and the anti-projected tensor takes the form $%
\bar DM=\bar O\left( \nabla M\right) =\left( I-O\right) \left( \nabla
M\right) $. For example, if $M\in T\left[ ^{VH}{}_{HV}\right] _P$, then $M$
obeys the projection identity%
$$
V^\alpha \!_\sigma {}H^\beta \!_\rho {}H^\tau \!_\mu {}M^{\sigma \rho
}\!_{\tau \nu }{}V^\nu \!_\gamma {}=M^{\alpha \beta }\!_{\mu \gamma }{}
$$
and its projected covariant derivative $DM$ has components
\begin{equation}
D_\delta {}M^{\alpha \beta }\!_{\mu \gamma
}{}=V^{\,\alpha }\!_\sigma {}H^\beta \!_\rho {}H^{\,\tau }\!_\mu {}\nabla
_\delta {}M^{\sigma \rho }\!_{\tau \nu }V^{\,\nu }\!_\gamma {}
\label{prjctd.deriv.2}
\end{equation}
while its antiprojected covariant derivative $\bar DM$ has components%
$$
\bar D_\delta M^{\alpha \beta }\!_{\mu \gamma }{}=\nabla _\delta {}M^{\alpha
\beta }\!_{\mu \gamma }-V^{\,\alpha }\!_\sigma {}H^\beta \!_\rho {}H^{\,\tau
}\!_\mu {}\nabla _\delta {}M^{\sigma \rho }\!_{\tau \nu }V^{\,\nu }\!_\gamma
$$

A point about notation: The projected and anti-projected covariant
derivatives $D,~\bar D$ carry no indication of what projection tensor fields
are to be used for evaluating them. They act only on tensor fields which are
identified as belonging to particular projected subspaces and {\em inherit}
the information about what projections to make from the object which is
being differentiated. One consequence of this inheritance property is that
these derivatives obey the product rule only for products of fully projected
tensors.

{}From the decomposition of the projection gradient given in Eq. (\ref
{prjctn.grad.decomp}) and the projection identities obeyed by the projection
curvatures given in Eqs. (\ref{h.prjctn.ids},\ref{hT.prjctn.ids}) it is
easy to show that the projected derivatives give zero when they act on the
projection tensor fields themselves.
\begin{equation}
D_\delta {}H^\alpha {}_\beta {}=0,\qquad D_\delta
{}V^\alpha {}_\beta {}=0.
\label{prjctn.compat}
\end{equation}

To relate the projected covariant derivative to the ordinary covariant
derivative, just take the covariant derivative of the projection identity
which the fully projected tensor obeys. For the simple case of a projected
vector field $v\left( P\right) \in HT_P$, the projection identity is%
$$
v^\alpha =H^\alpha \!_\rho {}v^\rho
$$
which becomes%
$$
\nabla _\delta \,v^\alpha =\left( \nabla _\delta {}H^\alpha \!_\rho
{}\right) v^\rho +H^\alpha \!_\rho {}\nabla _\delta \,v^\rho .
$$
Use the decomposition of the projection gradient eq.(\ref{prjctn.grad.decomp}%
) , the definition of the projected derivative, eq.(\ref{prjctd.deriv.1}),
and the projection identities obeyed by the projection curvature tensors
(Eqs. (\ref{h.prjctn.ids},\ref{hT.prjctn.ids}))together with the one
obeyed by $v$ to obtain the relation
\begin{equation}
\nabla _\delta {}v^\alpha {}=D_\delta {}v^\alpha
+\left( h_H{}^\alpha {}_{\rho \delta }{}-h_V^T{}_\rho {}^\alpha \!_\delta
{}\right) v^\rho .
\label{v.deriv.decomp}
\end{equation}
For a projected one-form field $\phi \left( P\right) \in H^{*}\hat T_P$ the
same procedure yields
\begin{equation}
\nabla _\delta {}\phi _\beta {}=D_\delta {}\phi _\beta
{}+\phi _\rho {}\left( h_H^T{}_\beta {}^\rho \!_\delta {}-h_V{}^\rho
{}_{\beta \delta }{}\right)
\label{f.deriv.decomp}
\end{equation}
For vectors and forms obeying projection identities with the complementary
projection $V$, just take the complements of these results.

In terms of the anti-projected covariant derivative, these last results take
the form:
\begin{equation}
\bar D_\delta v^\alpha =\left( h_H{}^\alpha {}_{\rho
\delta }{}-h_V^T{}_\rho {}^\alpha \!_\delta {}\right) v^\rho ,\qquad \bar
D_\delta \phi _\beta =\phi _\rho {}\left( h_H^T{}_\beta {}^\rho \!_\delta
{}-h_V{}^\rho {}_{\beta \delta }{}\right)
\label{v.aderiv.decomp}
\end{equation}
It is evident that the anti-projected derivative ignores how the vector
field $v$ changes within the subspaces $HT_P$. More generally, if a
gauge-field assigns a transformation $\Lambda \left( P\right)
:HT_P\rightarrow HT_P$ to each point $P$ of a manifold, and $v\left(
P\right) \in HT_P$ then the antiprojected derivative has the property:%
$$
\bar D\left( \Lambda v\right) =\Lambda \left( \bar Dv\right)
$$
Thus, the gauge-transformation field $\Lambda $ does not get differentiated
and acts purely locally. This gauge-locality property of the anti-projected
covariant derivative makes it a natural ingredient in any theory which
possesses a gauge group.

\subsubsection{Intrinsic and Extrinsic Projected and Anti-projected
Covariant Derivatives}

The projected and anti-projected covariant derivatives are not quite what I
want because they are not fully projected. The desired fully projected
objects are the {\em intrinsic projected covariant derivative} $D_H\,v$ with
components%
$$
D_{H\,\delta }\,v^\alpha {}=H^\rho \!_\delta {}D_\rho \,v^\alpha {}
$$
the {\em extrinsic projected covariant derivative} $D_V\,v$ with components%
$$
D_{V\,\delta }\,v^\alpha {}=V^\rho \!_\delta {}D_\rho \,v^\alpha {}
$$
the {\em intrinsic anti-projected covariant derivative} $\bar D_H\,v$ with
components%
$$
\bar D_{H\,\delta }\,v^\alpha {}=H^\rho \!_\delta {}\bar D_\rho \,v^\alpha
{}
$$
and the {\em extrinsic anti-projected covariant derivative} $\bar D_V\,v$
with components%
$$
\bar D_{V\,\delta }\,v^\alpha {}=V^\rho \!_\delta {}\bar D_\rho \,v^\alpha
{}
$$
Projecting Eqs. (\ref{v.deriv.decomp},\ref{f.deriv.decomp}) and using
the projection identities obeyed by the projection curvatures (equations (%
\ref{h.prjctn.ids},\ref{hT.prjctn.ids})) yields the following full
decomposition of the covariant derivative:
\begin{equation}
\begin{array}{c}
H^\rho \!_\delta {}\nabla _\rho v^\alpha {}=D_{H\,\delta }\,v^\alpha
{}+\,v^\rho {}h_H{}^\alpha {}_{\rho \delta }{} \\
V^\rho \!_\delta {}\nabla _\rho v^\alpha {}=D_{V\,\delta }\,v^\alpha
{}-v^\rho {}h_V^T{}_\rho {}^\alpha \!_\delta {} \\
H^\rho \!_\delta {}\nabla _\rho \phi _\beta {}=D_{H\,\delta }\,\phi _\beta
{}+\phi _\rho {}h_H^T{}_\beta {}^\rho \!_\delta {} \\
V^\rho \!_\delta {}\nabla _\rho \phi _\beta {}=D_{V\,\delta }\,\phi _\beta
{}-\phi _\rho {}h_V{}^\rho {}_{\beta \delta }{}
\end{array}
\label{cov.deriv.decomp}
\end{equation}
These expressions refer to a vector field $v$ with $v\left( P\right) \in
HT_P $ and a form-field $\phi $ with $\phi \left( P\right) \in H^{*}\hat T_P$%
. The expressions for $v\left( P\right) \in VT_P$ and $\phi \left( P\right)
\in V^{*}\hat T_P$ can be obtained by taking the complements --- exchange $H$
and $V$ everywhere. The net result of these exchanges is just to exchange $%
h_H{}_{\rho \delta }\!^\alpha {}$ and $-h_H^T{}^\alpha \!_{\delta \rho }{}$
and similarly exchange $h_V$ and $-h_V^T$ in the above expressions.

The anti-projected versions of these results are just%
$$
\begin{array}{c}
\bar D_{H\,\delta }\,v^\alpha {}=\,v^\rho {}h_H{}^\alpha {}_{\rho \delta
}{},\qquad \bar D_{V\,\delta }\,v^\alpha {}=-v^\rho {}h_V^T{}_\rho {}^\alpha
\!_\delta {} \\
\bar D_{H\,\delta }\,\phi _\beta {}=\phi _\rho {}h_H^T{}_\beta {}^\rho
\!_\delta {},\qquad \bar D_{V\,\delta }\,\phi _\beta {}=-\phi _\rho
{}h_V{}^\rho {}_{\beta \delta }{}
\end{array}
$$

\subsubsection{Fully Projected Decompositions of Covariant Derivatives}

For a general, fully projected tensor, the decomposition of the covariant
derivative has the same structure as the vector and form decomposition shown
above, but with a correction term for each index of the tensor. For example,
if $M\in T\left[ ^{V~H}{}_{H~V}\right] _P$ then the covariant derivative has
the decomposition%
$$
\begin{array}{c}
H^\rho \!_\delta {}\nabla _\rho \,{}M^{\alpha \beta }\!_{\mu \nu
}{}=D_{H\delta }\,{}M^{\alpha \beta }\!_{\mu \nu }{}-M^{\rho \beta }\!_{\mu
\nu }{}h_H^T{}_\rho {}^\alpha \!_\delta {}+M^{\alpha \rho }\!_{\mu \nu
}{}h_H{}^\beta {}_{\rho \delta }{} \\
+M^{\alpha \beta }\!_{\rho \nu }{}h_H^T{}_\mu {}^\rho \!_\delta {}-M^{\alpha
\beta }\!_{\mu \rho }{}h_H{}^\rho {}_{\nu \delta }{}
\end{array}
$$
$$
\begin{array}{c}
\begin{array}{c}
V^\rho \!_\delta {}\nabla _\rho \,{}M^{\alpha \beta }\!_{\mu \nu
}{}=D_{V\delta }\,{}M^{\alpha \beta }\!_{\mu \nu }{}+{}M^{\rho \beta
}\!_{\mu \nu }{}h_V{}^\alpha {}_{\rho \delta }{}-M^{\alpha \rho }\!_{\mu \nu
}{}h_V^T{}_\rho {}^\alpha \!_\delta {} \\
-M^{\alpha \beta }\!_{\rho \nu }{}h_V{}^\rho {}_{\nu \delta }{}+M^{\alpha
\beta }\!_{\mu \rho }{}h_V^T{}_\nu {}^\rho \!_\delta {}
\end{array}
\end{array}
$$
The projection correction terms in this sort of decomposition can be written
with the help of just three rules:

\begin{itemize}
\item  (1) Contract each tensor index in turn with the first or second index
of one of the four projection curvature tensors $h_H,h_H^T,h_V,h_V^T$. Set
the last index on the projection curvature equal to the differentiating
index. Set the remaining index equal to the tensor index which is being
corrected.

\item  (2) Choose the projection curvature tensor whose indexes are in the
right positions (up or down) and have the correct projection properties to
yield a consistent non-zero term --- For each index, only one choice will
work. (Do not raise or lower indexes.)

\item  (3) When the corrected tensor index obeys a projection identity
complementary to the one obeyed by the differentiating index, the correction
term has a minus sign. Otherwise it has a plus sign.
\end{itemize}

At some risk of taking excessive poetic license, I will refer to these rules
as the {\em generalized Gauss-Weingarten Relations.} When all of the
differences in notation and point of view have been swept away, these rules
do the essential job of the Gauss-Weingarten relations: They relate the full
connection to the projected connection.\cite{Gauss-Weingarten}

One consequence of these rules is that when the differentiating index has
been projected with $H$, then the correction terms can only be constructed
from $h_H,h_H^T.$ Similarly when the differentiating index has been
projected with $V$, then the correction terms can only be constructed from $%
h_V,h_V^T$. Note the projection identities obeyed by the projection
curvatures (Eqs. (\ref{hT.prjctn.ids},\ref{hcomp.prjctn.ids})). A
useful consequence of these identities is that the correction term which is
associated with a given tensor index is always projected in a manner {\em %
complementary} to that of the original tensor index. This consequence can be
used as a consistency check. It also means that the correction terms which
are associated with contracted indexes often vanish.

A final point about notation: The operators $D_H,D_V,\bar D_H,\bar D_V$
specify the projection which is to be performed on the differentiating index
of the tensors which they produce. However, just like the projected
derivative $D$, they do not specify the projections which are to be
performed on the other indexes. Those projections are ''inherited'' from the
tensor fields which are being differentiated. Thus, although the situations
which have been considered so far involve only combinations of a single
projection tensor field, $H,$ and its complement, $V$, there will be cases
(such as null projections) where more than one projection tensor field is
present. In those cases, one may have an operation $D_HT$ which projects the
indexes inherited from the tensor $T$ with a projection tensor field which
is neither $H$ nor its complement $V$. One may also have an object which
belongs to more than one projection subspace so that it needs to be assigned
a ''home subspace'' for its projected derivative operators to be defined.

\section{Geometrical Structure Decompositions}
\label{gstructure}
\subsection{Metricity}

When a form-metric with components $g^{\mu \nu }$ exists, the metricity
tensor is just the covariant derivative of the metric which can be
decomposed by the rules of the previous section. Here, I will take $g^{\mu
\nu }$ to be an {\em arbitrary tensor field} which need not have all of the
properties which we usually associate with a metric. For example, the
vector-metric $g_{\mu \nu }$ may not exist. First, decompose this 'metric'
into fully projected tensors:%
$$
g^{XY}{}^{\mu \nu }{}=g\left[ ^{XY}\right] {}^{\mu \nu }{}=g^{\rho \sigma
}X^\mu {}_\rho {}Y^\nu {}_\sigma {}
$$
$$
g^{\mu \nu }{}=\,g^{HH}{}^{\mu \nu }{}+\,g^{HV}{}^{\mu \nu
}{}+\,g^{VH}{}^{\mu \nu }{}+\,g^{VV}{}^{\mu \nu }{}
$$
where the projection labels $X,Y$ stand for either $H$ or $V$ and express
the metricity, $Q^{\mu \nu }{}_\rho {}=-\nabla _\rho {}g^{\mu \nu }{}$, in
terms of these:%
$$
\begin{array}{c}
-Q^{\mu \nu }{}_\rho {}H^\rho {}_\delta {}=D_{H\delta }{}g^{HH}{}^{\mu \nu
}{}+g^{HH\,\rho \nu }{}h_H{}^\mu {}_{\rho \delta }{}+g^{HH\,\mu \rho
}{}h_H{}^\nu {}_{\rho \delta }{} \\
+D_{H\delta }\,g^{HV}{}^{\mu \nu }{}+g^{HV}{}^{\rho \nu }{}h_H{}^\mu
{}_{\rho \delta }{}-g^{HV}{}^{\mu \rho }{}h_H^T{}_\rho {}^\nu {}_\delta {}
\\
+D_{H\delta }\,g^{VH}{}^{\mu \nu }{}-g^{VH}{}^{\rho \nu }{}h_H^T{}_\rho
{}^\mu {}_\delta {}+g^{VH}{}^{\mu \rho }{}h_H{}^\nu {}_{\rho \delta }{} \\
+D_{H\delta }\,g^{VV}{}^{\mu \nu }{}-g^{VV}{}^{\rho \nu }{}h_H^T{}_\rho
{}^\mu {}_\delta {}-g^{VV}{}^{\mu \rho }{}h_H^T{}_\rho {}^\nu {}_\delta {}
\end{array}
$$
The decomposition of the projection $Q^{\mu \nu }{}_\rho {}V^\rho {}_\delta
{}$ is then obtained by taking the complement of this result.

When $H$ projects onto surfaces, the quantity $D_{H\delta }{}g^{HH}{}^{\mu
\nu }{}$ is the metricity of the intrinsic geometry on those surfaces. In
general, I define the {\em intrinsic metricity} to be the projected
intrinsic derivative
$$
Q_H^{HH}{}^{\mu \nu }{}_\delta {}=-D_{H\delta }{}g^{HH}{}^{\mu \nu }{}
$$
and, to complete the decomposition of the metricity, I define {\em %
cross-projected metricities}
$$
Q_H^{XY}{}^{\mu \nu }{}_\delta {}=-D_{H\delta }{}g^{XY}{}^{\mu \nu }{}
$$
as well as the complements of these objects. My earlier caution (see Section
\ref{prjctd.antiprjctd}) about the
product rule for projected derivatives of tensor products comes into play
here. If the connection is metric compatible, one might suspect that the
intrinsic and cross-projected metricities would automatically vanish as a
consequence of Eq. (\ref{prjctn.compat}). As is shown next, they do not
necessarily vanish.

Project out the different components of the metricity:
\begin{equation}
Q\left[ ^{HH}{}_H\right] {}^{\mu \nu }{}_\delta
{}=Q_H^{HH}{}^{\mu \nu }{}_\delta {}+g^{HV}{}^{\mu \rho }{}h_H^T{}_\rho
{}^\nu {}_\delta {}+g^{VH}{}^{\rho \nu }{}h_H^T{}_\rho {}^\mu {}_\delta {}
\label{intr.metricity}
\end{equation}
\begin{equation}
Q\left[ ^{HH}{}_V\right] {}^{\mu \nu }{}_\delta
{}=Q_V^{HH}{}^{\mu \nu }{}_\delta {}-g^{HV}{}^{\mu \rho }{}h_V{}^\nu
{}_{\rho \delta }{}-g^{VH}{}^{\rho \nu }{}h_V{}^\mu {}_{\rho \delta }{}
\label{fermi.metricity}
\end{equation}
\begin{equation}
Q\left[ ^{HV}{}_H\right] {}^{\mu \nu }{}_\delta
{}=Q_H^{HV}{}^{\mu \nu }{}_\delta {}-g^{HH}{}^{\mu \rho }{}h_H{}^\nu
{}_{\rho \delta }{}+g^{VV}{}^{\rho \nu }{}h_H^T{}_\rho {}^\mu {}_\delta {}
\label{hT-h.relation}
\end{equation}
\begin{equation}
Q\left[ ^{HV}{}_V\right] {}^{\mu \nu }{}_\delta
{}=Q_V^{HV}{}^{\mu \nu }{}_\delta {}+g^{HH}{}^{\mu \rho }{}h_V^T{}_\rho
{}^\nu {}_\delta {}-g^{VV}{}^{\rho \nu }{}h_V{}^\mu {}_{\rho \delta }{}
\label{hTV-h.relation}
\end{equation}
Ordinarily the connection is metric compatible so that the metricity tensor
vanishes and all of the above equations have zero on their left-hand sides.
For a normal projection, the cross-projected metric $g^{HV}$ is zero. When
both these conditions hold, the above equations simply say that both the
intrinsic and the cross-projected metricity tensors vanish and the two types
of projection curvature are the same: $h=h^T$. If, however, the connection
is metric compatible but the projection is not normal, the above equations
yield interesting results including:%
$$
Q_H^{HH}{}^{\mu \nu }{}_\delta {}=-g^{HV}{}^{\mu \rho }{}h_H^T{}_\rho {}^\nu
{}_\delta {}-g^{VH}{}^{\rho \nu }{}h_H^T{}_\rho {}^\mu {}_\delta {}.
$$
The intrinsic and cross-projected metricities {\em do not necessarily vanish}
for non-normal projection tensor fields even if the connection is metric
compatible.

\subsection{Torsion}

\subsubsection{Definition and Projection}

The torsion tensor $S^\rho {}_{\mu \nu }{}$ is defined by the relation
\begin{equation}
\left[ \nabla _\nu {},\nabla _\mu {}\right] \phi =S^\rho
{}_{\mu \nu }{}\nabla _\rho {}\phi .
\label{torsion.def}
\end{equation}
for any function $\phi $ on the manifold. To decompose this relation into
fully projected parts, begin by decomposing the gradient%
$$
\nabla _\alpha {}\phi =D_{H\alpha }{}\phi +D_{V\alpha }{}\phi
$$
so that the definition of torsion becomes%
$$
\nabla _\nu {}D_{H\mu }{}\phi +\nabla _\nu {}D_{V\mu }{}\phi -\nabla _\mu
{}D_{H\nu }{}\phi -\nabla _\mu {}D_{V\nu }{}\phi =S^\rho {}_{\mu \nu
}{}\nabla _\rho {}\phi .
$$
Project the two free indexes $\mu ,\nu $ with $H$ and use the definition of
the intrinsic projected derivative as well as Eq.(\ref{cov.deriv.decomp}%
) to obtain the $HH$-projection
\begin{equation}
\left[ D_{H\nu }{},D_{H\mu }{}\right] \phi =2\omega
_H{}^\rho {}_{\mu \nu }{}D_{V\rho }{}\phi +S\left[ ^I{}_{H~H}\right] {}^\rho
{}_{\mu \nu }{}\nabla _\rho {}\phi
\label{int.int.derivs}
\end{equation}
of the torsion definition. Project one free index with $H$ and the other
with $V$ and proceed as before to obtain the $HV$-projection
\begin{equation}
\begin{array}{c}
\left[ D_{V\,\nu }{},D_{H\mu }{}\right] \phi -h_H^T{}_\nu {}^\rho {}_\mu
{}D_{H\rho }{}\phi +h_V^T{}_\mu {}^\rho {}_\nu {}D_{V\rho }{}\phi \\
=S\left[ ^H{}_{H~V}\right] {}^\rho {}_{\mu \nu }{}D_{H\rho }{}\phi +S\left[
^V{}_{H~V}\right] {}^\rho {}_{\mu \nu }{}D_{V\rho }{}\phi
\end{array}
\label{int.ext.derivs}
\end{equation}
of the torsion definition.

\subsubsection{Intrinsic and Cross Torsions}

When $H$ projects onto surfaces, the quantity $\left[ D_{H\nu },D_{H\mu
}\right] \phi $ is simply related to the torsion of the intrinsic geometry
on those surfaces. In general, define the {\em intrinsic and cross torsion}
tensors $S_{HH}^H,S_{HH}^V,S_{HV}^H,S_{HV}^V$ and their complements by%
$$
\left[ D_{Y\,\nu }{},D_{X\mu }{}\right] \phi =S_{XY}^H{}^\rho {}_{\mu \nu
}{}D_{H\rho }{}\phi +S_{XY}^V{}^\rho {}_{\mu \nu }{}D_{V\rho }{}\phi
$$
{}From the decompositions (Equations
(\ref{int.int.derivs},\ref{int.ext.derivs}%
))above,
$$
\begin{array}{c}
S\left[ ^H{}_{H~H}\right] {}^\rho {}_{\mu \nu }{}D_{H\rho }{}\phi +S\left[
^V{}_{H~H}\right] {}^\rho {}_{\mu \nu }{}D_{V\rho }{}\phi +2\omega _H{}^\rho
{}_{\mu \nu }{}D_{V\rho }{}\phi \\
=S_{HH}^H{}^\rho {}_{\mu \nu }{}D_{H\rho }{}\phi +S_{HH}^V{}^\rho {}_{\mu
\nu }{}D_{V\rho }{}\phi
\end{array}
$$
$$
\begin{array}{c}
S\left[ ^H{}_{H~V}\right] {}^\rho {}_{\mu \nu }{}D_{H\rho }{}\phi +S\left[
^V{}_{H~V}\right] {}^\rho {}_{\mu \nu }{}D_{V\rho }{}\phi +h_H^T{}_\nu
{}^\rho {}_\mu {}D_{H\rho }{}\phi -h_V^T{}_\mu {}^\rho {}_\nu {}D_{V\rho
}{}\phi \\
=S_{HV}^H{}^\rho {}_{\mu \nu }{}D_{H\rho }{}\phi +S_{HV}^V{}^\rho {}_{\mu
\nu }{}D_{V\rho }{}\phi
\end{array}
$$
which give the decompositions:
\begin{equation}
S\left[ ^H{}_{H~H}\right] {}^\rho {}_{\mu \nu
}{}=S_{HH}^H{}^\rho {}_{\mu \nu }{}
\end{equation}
\begin{equation}
\label{vhh.tor.def}S\left[ ^V{}_{H~H}\right] {}^\rho {}_{\mu \nu
}{}=S_{HH}^V{}^\rho {}_{\mu \nu }{}-2\omega _H{}^\rho {}_{\mu \nu }{}
\label{int.tor.def}
\end{equation}
\begin{equation}
S\left[ ^H{}_{H~V}\right] {}^\rho {}_{\mu \nu
}{}=S_{HV}^H{}^\rho {}_{\mu \nu }{}-h_H^T{}_\nu {}^\rho {}_\mu {}
\label{hhv.tor.def}
\end{equation}
and their complements.

\subsubsection{Surface Formation: Frobenius Theorem}

When a projection tensor field $H$ yields projected tangent spaces $HT_P$
which are the tangent spaces to a system of submanifolds, each submanifold
has its own, fully self-contained {\em intrinsic geometry}.\cite
{intrinsic.note} When these intrinsic geometries exist, they provide
powerful computational tools and important insights as in the 3+1
decomposition of the initial value problem of general relativity for
example. Ordinarily, however, the subspaces and intrinsic derivative
operations associated with a given projection tensor field $H$ do not form
fully self-contained intrinsic geometries. What conditions on a projection
tensor field are sufficient for intrinsic geometries to exist?

In an adapted frame, the $HH$-projection of the torsion definition (Eq.
(\ref{int.int.derivs}) above) takes the form
\begin{equation}
\left\{ \left[ e_n{},e_m{}\right] -\left( 2\Gamma
^r{}_{\left[ mn\right] }+S_{HH}^H{}^r{}_{mn}{}\right)
e_r{}-S_{HH}^V{}^R{}_{mn}{}e_R{}\right\} \phi =0.
\label{hh.torsion.adapt}
\end{equation}
while the $HV$-projection is%
$$
\left\{ \left[ e_E{},e_d{}\right] -\left( \Gamma
^r{}_{dE}{}+S_{HV}^H{}^r{}_{dE}{}\right) e_r{}+\left( \Gamma
^R{}_{Ed}{}-S_{HV}^{V\,}{}^R{}_{dE}{}\right) e_R\right\} {}\phi =0
$$
The \thinspace $VV$-projection can be obtained by taking the complement of
Eq. (\ref{hh.torsion.adapt}) -- replace $H$ by $V$ and switch upper and
lower case indexes everywhere. Notice how these results simplify when
expressed in terms of the cross-torsion tensor components given in equations
(\ref{vhh.tor.def},\ref{hhv.tor.def}) above.

The $HH$-projection of the torsion tensor definition given by Eq. (\ref
{hh.torsion.adapt}) above provides the needed relation. This result shows
that two vector fields with values in the subspaces $HT_P$ have a commutator
which lies in the same subspaces if and only if the cross-projected torsion
tensor $S_{HH}^{V\,}$ is zero. The Frobenius theorem then guarantees that
the subspaces are tangent to a system of submanifolds. Thus, the vanishing
of the tensor $S_{HH}^{V\,}$ is a necessary and sufficient condition for a
projection tensor field to yield subspaces which are tangent to submanifolds.

In a torsion-free geometry, the cross-projected torsion $S_{HH}^{V\,}$ is
not always zero. From Equation (\ref{vhh.tor.def}) it is related to the
twist or vorticity tensor by $S_{HH}^V{}^\rho {}_{\mu \nu }{}=2\omega
_H{}^\rho {}_{\mu \nu }{}$. For this reason, I will call this particular
cross-torsion tensor, the {\em generalized twist} of a projection tensor
field.

\subsection{Riemann Curvature}

\subsubsection{Definition and Projection}

The curvature tensor is defined by the equation
\begin{equation}
v^\rho {}R_\rho {}^\gamma {}_{\alpha \beta }{}=\left( \left[
\nabla _\beta {},\nabla _\alpha {}\right] -S^\rho {}_{\alpha \beta }{}\nabla
_\rho {}\right) v^\gamma {}.
\label{curv.def}
\end{equation}
By letting the torsion definition act on the function $\phi _\gamma
{}v^\gamma {}$ one finds that this definition may be restated in terms of
derivatives acting on one-forms:
\begin{equation}
\left( \left[ \nabla _\beta {},\nabla _\alpha {}\right]
-S^\rho {}_{\alpha \beta }{}\nabla _\rho {}\right) \phi _\gamma {}=-R_\gamma
{}^\rho {}_{\alpha \beta }{}\phi _\rho {}.
\label{alt.curv.def}
\end{equation}
Either form of the definition can be decomposed by using the rules given in
section \ref{prjctd.cnctn}. Start with a vector $v$ such that $Hv=v$ and
decompose the first derivative
$$
\nabla _\alpha {}v^\gamma =D_{H\alpha }{}v^\gamma {}+v^\rho {}h_H{}^\gamma
{}_{\rho \alpha }{}+D_{V\alpha }{}v^\gamma {}-v^\rho {}h_V^T{}_\rho
{}^\gamma {}_\alpha {}
$$
and then, in a straightforward calculation, the second derivative, $\nabla
_\beta {}\nabla _\alpha {}v^\gamma {}$. Use this result to evaluate Eq.
(\ref{curv.def}) and form all of its independent projections. The results
become simple and symmetrical when they are expressed in terms of the
intrinsic and cross torsion tensors.
\begin{equation}
\begin{array}{c}
v^\rho {}R\left[ _H{}^H{}_{HH}\right] {}_\rho {}^\gamma {}_{\alpha \beta
}{}=\left( \left[ D_{H\beta }{},D_{H\alpha }{}\right] -S_{HH}^H{}^\rho
{}_{\alpha \beta }{}D_{H\rho }{}-S_{HH}^V{}^\rho {}_{\alpha \beta
}{}D_{V\rho }{}\right) v^\gamma {} \\
+v^\rho {}\left( -h_H^T{}_\sigma {}^\gamma {}_\beta {}h_H{}^\sigma {}_{\rho
\alpha }{}+h_H^T{}_\sigma {}^\gamma {}_\alpha {}h_H{}^\sigma {}_{\rho \beta
}{}\right)
\end{array}
\label{curv.hhhh}
\end{equation}
\begin{equation}
\begin{array}{c}
v^\rho {}R\left[ _H{}^H{}_{HV}\right] {}_\rho {}^\gamma {}_{\alpha \beta
}{}=\left( \left[ D_{V\beta }{},D_{H\alpha }{}\right] -S_{HV}^H{}^\rho
{}_{\alpha \beta }{}D_{H\rho }{}-S_{HV}^V{}^\rho {}_{\alpha \beta
}{}D_{V\rho }{}\right) v^\gamma {} \\
+v^\rho {}\left( h_H{}^\sigma {}_{\rho \alpha }{}h_V{}^\gamma {}_{\sigma
\beta }{}-h_H^T{}_\sigma {}^\gamma {}_\alpha {}h_V^T{}_\rho {}^\sigma
{}_\beta {}\right)
\end{array}
\label{curv.hhhv}
\end{equation}
\begin{equation}
\begin{array}{c}
v^\rho {}R\left[ _H{}^H{}_{VV}\right] {}_\rho {}^\gamma {}_{\alpha \beta
}{}=\left( \left[ D_{V\beta }{},D_{V\alpha }{}\right] -S_{VV}^H{}^\rho
{}_{\alpha \beta }{}D_{H\rho }{}-S_{VV}^V{}^\rho {}_{\alpha \beta
}{}D_{V\rho }{}\right) v^\gamma {} \\
+v^\rho {}\left( h_V{}^\gamma {}_{\sigma \alpha }{}h_V^T{}_\rho {}^\sigma
{}_\beta {}-h_V{}^\gamma {}_{\sigma \beta }{}h_V^T{}_\rho {}^\sigma
{}_\alpha {}\right)
\end{array}
\label{curv.hhvv}
\end{equation}
\begin{equation}
\begin{array}{c}
v^\rho {}R\left[ _H{}^V{}_{HH}\right] {}_\rho {}^\gamma {}_{\alpha \beta }{}
\\
=v^\rho {}\left( D_{H\beta }{}h_H{}^\gamma {}_{\rho \alpha }{}-D_{H\alpha
}{}h_H{}^\gamma {}_{\rho \beta }{}-h_H{}^\gamma {}_{\rho \sigma
}{}S_{HH}^H{}^\sigma {}_{\alpha \beta }{}+h_V^T{}_\rho {}^\gamma {}_\sigma
{}S_{HH}^V{}^\sigma {}_{\alpha \beta }{}\right)
\end{array}
\label{curv.hvhh}
\end{equation}
\begin{equation}
\begin{array}{c}
v^\rho {}R\left[ _H{}^V{}_{HV}\right] {}_\rho {}^\gamma {}_{\alpha \beta }{}
\\
=v^\rho {}\left( D_{V\beta }{}h_H{}^\gamma {}_{\rho \alpha }{}-D_{H\alpha
}{}h_V^T{}_\rho {}^\gamma {}_\beta {}-h_H{}^\gamma {}_{\rho \sigma
}{}S_{HV}^H{}^\sigma {}_{\alpha \beta }{}+h_V^T{}_\rho {}^\gamma {}_\sigma
{}S_{HV}^V{}^\sigma {}_{\alpha \beta }{}\right)
\end{array}
\label{curv.hvhv}
\end{equation}
\begin{equation}
\begin{array}{c}
v^\rho {}R\left[ _H{}^V{}_{VV}\right] {}_\rho {}^\gamma {}_{\alpha \beta }{}
\\
=v^\rho {}\left( -D_{V\beta }{}h_V^T{}_\rho {}^\gamma {}_\alpha
{}+D_{V\alpha }{}h_V^T{}_\rho {}^\gamma {}_\beta {}-h_H{}^\gamma {}_{\rho
\sigma }{}S_{VV}^H{}^\sigma {}_{\alpha \beta }{}+h_V^T{}_\rho {}^\gamma
{}_\sigma {}S_{VV}^V{}^\sigma {}_{\alpha \beta }{}\right)
\end{array}
\label{curv.hvvv}
\end{equation}
The rest of the projections of the curvature tensor obey equations which are
the complements of these.

\subsubsection{Intrinsic and Cross-projected Curvature Tensors}

Because each of these equations is an identity, the expressions on their
right-hand sides must be strictly local in the vector field $v$. Thus, the
combinations of intrinsic and extrinsic projected derivatives which appear
lead to the identification of several new tensors. Six new tensors are
defined as follows: For any vector field $v$ such that $Zv=v$,
\begin{equation}
v^\rho {}R_{XY}^Z{}_\rho {}^\gamma {}_{\alpha \beta
}{}=\left( \left[ D_{Y\beta }{},D_{X\alpha }{}\right] -S_{XY}^H{}^\rho
{}_{\alpha \beta }{}D_{H\rho }{}-S_{XY}^V{}^\rho {}_{\alpha \beta
}{}D_{V\rho }{}\right) v^\gamma {}
\label{Rhhhh.def}
\end{equation}
where each of the projection labels $X,Y,Z$ can be either $H$ or $V$. Since
the right-hand side of each of these equations gives zero for a vector field
$v$ such that $\bar Zv=v$, it is clear that each of these tensors belongs to
a projected subspace: $R_{XY}^Z(P)\in T_P\left[ _Z{}^Z{}_{XY}\right] $.
Notice that the first two arguments of these tensors are always in the same
projected subspace. It is essential that the generalized commutation
operators which define these tensors map projected subspaces into themselves.

A notation which will be used later replaces the upper projection label on
the tensor $R_{XY}^Z$ by the product of two projection tensors. Thus, $%
R_{XY}^{Z\!W}$ is a tensor whose upper label is the result of the product $%
ZW $ when that product is a projection tensor. When $ZW$ is zero, the tensor
$R_{XY}^{Z\!W}$ is also zero. For example:%
$$
R_{HV}^{H\!H}{}=R_{HV}^H,\qquad R_{HV}^{V\!V}{}=R_{HV}^V,\qquad
R_{HV}^{H\!V}{}=0
$$
With this notation, $R_{XY}^{Z\!W}(P)\in T_P\left[ _Z{}^W{}_{XY}\right] $
and there is a (partly trivial) correspondence between the projection labels
and the indexes of the tensor.

Each of these tensors may also be defined for one-form fields in the same
way as the full curvature tensor. For example, $R_{VV}^H$ can be defined by
requiring for any $\eta $ with $\eta (P)$$\in H^{*}T_P$,
$$
-\eta _\rho {}R_{VV}^H{}_\gamma {}^\rho {}_{\alpha \beta }{}==\left( \left[
D_{V\beta }{},D_{V\alpha }{}\right] -S_{VV}^H{}^\rho {}_{\alpha \beta
}{}D_{H\rho }{}-S_{VV}^V{}^\rho {}_{\alpha \beta }{}D_{V\rho }{}\right) \eta
_\gamma {}.
$$
These one-form versions of the definitions can be obtained from the vector
forms by repeating the usual argument for the full curvature tensor --- Let
the torsion definition act on the function $\eta _\rho {}v^\rho {}$ where $%
\eta (P)$ and $v(P)$ are restricted to $H^{*}T_P$ and $HT_P$. Alternatively,
one can decompose the one-form version of the curvature definition given by
Eq. (\ref{alt.curv.def}).

Two of these new tensors are familiar: When the projection $H$ is
surface-forming, the generalized twist tensor $S_{HH}^V$ vanishes and $%
R_{HH}^H$ is clearly the intrinsic curvature tensor of the surface. In an
adapted frame, the components of $R_{HH}^H$ are given by the
familiar-looking expression
\begin{equation}
\begin{array}{c}
R_{HH}^H{}_r{}^c{}_{ab}{}=e_b{}\left( \Gamma ^c{}_{ra}{}\right) -e_a{}\left(
\Gamma ^c{}_{rb}{}\right) \\
+\Gamma ^s{}_{ra}{}\Gamma ^c{}_{sb}{}-\Gamma ^s{}_{rb}{}\Gamma
^c{}_{sa}{}-\Gamma ^c{}_{rs}{}\left( 2\Gamma ^s{}_{\left[ ab\right]
}{}-S^{\,s}{}_{ab}{}\right) -S^{\,S}{}_{ab}{}\Gamma ^c{}_{rS}{}
\end{array}
\label{Rhhhh.adapt}
\end{equation}
For the general case, we define this tensor to be the {\em intrinsic
curvature tensor} of the projection tensor field $H$. Similarly, $R_{VV}^V$
is the intrinsic curvature of the projection tensor field $V$ and has an
adapted frame expression which is the complement of Eq. (\ref
{Rhhhh.adapt}).

For non-surface-forming projection-tensor fields, the intrinsic curvature
tensor is still an object which has been seen before, although not in the
generality which is presented here. It was developed by MacCallum in the
context of three-dimensional projections into the rest-frame subspaces of a
fluid in spacetime.\cite{intrinsic.geom.a,tcong.q.geom} Collins and Szafron
developed a scheme for classifying fluid-containing spacetimes by their
intrinsic curvature tensors\cite{intrinsic.geom,intrinsic.geom.I}, and
showed that the Szekeres solutions of Einstein's equations are examples of a
restricted class in this scheme.\cite{intrinsic.geom.II}

The remaining four tensors, $R_{HV}^H,R_{VV}^H,R_{VH}^V,R_{HH}^V$ are not so
familiar. They are essentially the commutators of intrinsic and extrinsic
derivatives. I will call them {\em cross-projected curvature tensors}. In an
adapted frame, these cross-projected curvature tensors have the expressions:%
$$
\begin{array}{c}
R_{HV}^H{}_r{}^a{}_{dE}=e_E{}\left( \Gamma ^a{}_{rd}{}\right) -e_d{}\left(
\Gamma ^a{}_{rE}{}\right) +\Gamma ^s{}_{rd}{}\Gamma ^a{}_{sE}{}-\Gamma
^s{}_{rE}{}\Gamma ^a{}_{sd}{} \\
-\left( \Gamma ^s{}_{dE}{}+h_H^T{}_E{}^s{}_d{}+S^{\,s}{}_{dE}{}\right)
\Gamma ^a{}_{rs}{}+\left( \Gamma
^R{}_{Ed}{}+h_V^T{}_d{}^R{}_E{}-S^{\,R}{}_{dE}{}\right) \Gamma ^a{}_{rR}{}
\end{array}
$$
$$
\begin{array}{c}
R_{VV}^H{}_r{}^c{}_{AB}=e_B{}\left( \Gamma ^c{}_{rA}{}\right) -e_A{}\left(
\Gamma ^c{}_{rB}{}\right) +\Gamma ^s{}_{rA}{}\Gamma ^c{}_{sB}{}-\Gamma
^s{}_{rB}{}\Gamma ^c{}_{sA}{} \\
-\left( 2\omega _V{}^s{}_{AB}{}+S^{\,s}{}_{AB}{}\right) \Gamma
^c{}_{rs}{}-\left( 2\Gamma ^R{}_{\left[ AB\right]
}{}+S^{\,R}{}_{AB}{}\right) \Gamma ^c{}_{rR}{}
\end{array}
$$
with $R_{VH}^V,R_{HH}^V$ given by the complement expressions. Notice that
the key ingredients in these tensors are precisely those mixed adapted-frame
connection coefficients $\Gamma ^c{}_{sB}{}$ which are not identified as
projection curvatures. Just as the intrinsic curvature $R_{HH}^H$ is the
simplest tensor which can be constructed from the intrinsic connection
coefficients $\Gamma ^a{}_{rd}{}$, the cross-curvatures are the simplest
tensors which can be constructed from the remaining connection coefficients.

The simple applications which are described in section \ref{applications}
reveal one reason that the cross-projected curvatures are unfamiliar: The
traditional applications of projection tensor methods all involve situations
where $VT_P$ is one-dimensional so that all of these cross-projected
curvatures either vanish identically or can be expressed in terms of the
projection curvatures via the Bianchi identities.

\subsubsection{Full Curvature Decomposition}

In terms of the intrinsic and cross curvature tensors, the decomposition of
the Riemannian curvature tensor becomes:

\begin{equation}
R\left[ _H{}^H{}_{HH}\right] {}_\rho \!^\gamma
\!_{\alpha \beta }=R_{HH\,}^H{}_\rho \!^\gamma \!_{\alpha \beta
}-h_H^T{}_\sigma {}^\gamma \!_\beta {}h_H{}^\sigma {}_{\rho \alpha
}\!+h_H^T{}_\sigma {}^\gamma \!_\alpha {}h_H{}^\sigma {}_{\rho \beta }\!
\label{R.hhhh.decomp}
\end{equation}
\begin{equation}
R\left[ _H{}^H{}_{HV}\right] {}_\rho \!^\gamma
\!_{\alpha \beta }=R_{HV}^H{}_\rho \!^\gamma \!_{\alpha \beta }+h_H{}^\sigma
{}_{\rho \alpha }\!h_V{}^\gamma {}_{\sigma \beta }\!-h_H^T{}_\sigma
{}^\gamma \!_\alpha {}h_V^T{}_\rho {}^\sigma \!_\beta {}
\label{R.hhhv.decomp}
\end{equation}
\begin{equation}
R\left[ _H{}^H{}_{VV}\right] {}_\rho \!^\gamma
\!_{\alpha \beta }=R_{VV}^H{}_\rho \!^\gamma \!_{\alpha \beta }+h_V{}^\gamma
{}_{\sigma \alpha }\!h_V^T{}_\rho {}^\sigma \!_\beta {}-h_V{}^\gamma
{}_{\sigma \beta }\!h_V^T{}_\rho {}^\sigma \!_\alpha {}
\label{R.hhvv.decomp}
\end{equation}
\begin{equation}
R\left[ _H{}^V{}_{HH}\right] {}_\rho \!^\gamma
\!_{\alpha \beta }=D_{H\beta }{}h_H{}^\gamma {}_{\rho \alpha }\!-D_{H\alpha
}{}h_H{}^\gamma {}_{\rho \beta }\!-h_H{}^\gamma {}_{\rho \sigma
}\!S_{HH}^H{}^\sigma \!_{\alpha \beta }{}+h_V^T{}_\rho {}^\gamma \!_\sigma
{}S_{HH}^V{}^\sigma \!_{\alpha \beta }{}
\label{R.hvhh.decomp}
\end{equation}
\begin{equation}
R\left[ _H{}^V{}_{HV}\right] {}_\rho \!^\gamma
\!_{\alpha \beta }=D_{V\beta }{}h_H{}^\gamma {}_{\rho \alpha }\!+D_{H\alpha
}{}h_V^T{}_\rho {}^\gamma \!_\beta {}-h_H{}^\gamma {}_{\rho \sigma
}\!S_{HV}^H{}^\sigma \!_{\alpha \beta }{}+h_V^T{}_\rho {}^\gamma \!_\sigma
{}S_{HV}^V{}^\sigma \!_{\alpha \beta }{}
\label{R.hvhv.decomp}
\end{equation}
\begin{equation}
R\left[ _H{}^V{}_{VV}\right] {}_\rho \!^\gamma
\!_{\alpha \beta }=-D_{V\beta }{}h_V^T{}_\rho {}^\gamma \!_\alpha
{}+D_{V\alpha }{}h_V^T{}_\rho {}^\gamma \!_\beta {}-h_H{}^\gamma {}_{\rho
\sigma }\!S_{VV}^H{}^\sigma \!_{\alpha \beta }{}+h_V^T{}_\rho {}^\gamma
\!_\sigma {}S_{VV}^V{}^\sigma \!_{\alpha \beta }{}
\label{R.hvvv.decomp}
\end{equation}

Some of these projections of the curvature tensor are familiar: Equations (%
\ref{R.hhhh.decomp},\ref{R.hvhh.decomp}) are the generalizations of the
Gauss-Codazzi relations. The complement of Equation (\ref{R.hvvv.decomp})
\begin{equation}
\label{R.vhhh.decomp}R\left[ _V{}^H{}_{HH}\right] {}_\rho \!^\gamma
\!_{\alpha \beta }=-D_{H\beta }{}h_H^T{}_\rho {}^\gamma \!_\alpha
{}+D_{H\alpha }{}h_H^T{}_\rho {}^\gamma \!_\beta {}-h_V{}^\gamma {}_{\rho
\sigma }\!S_{HH}^V{}^\sigma \!_{\alpha \beta }{}+h_H^T{}_\rho {}^\gamma
\!_\sigma {}S_{HH}^H{}^\sigma \!_{\alpha \beta }{}.
\end{equation}
is almost the same as Equation (\ref{R.hvhh.decomp}). When the projection is
normal and the connection is metric compatible, this last relation is just
Equation (\ref{R.hvhh.decomp}) with the first two indexes reversed. In
general, however, it is a necessary and independent addition to the
generalized Gauss-Codazzi relations.

Contractions and anti-symmetric parts of these projections provide other
useful results. From Eq. (\ref{R.hvhv.decomp}) one easily finds the
{\em divergence integrability} condition:
\begin{equation}
\begin{array}{c}
R\left[ _H{}^V{}_{HV}\right] {}_{[\rho }\!^\beta \!_{\alpha ]\beta
}{}=D_{V\beta }{}\omega _H{}^\beta {}_{\rho \alpha }\!+D_{H[\alpha }{}\theta
_V^T{}_{\rho ]}{} \\
+S_{VH}^H{}^\sigma \!_{\beta [\alpha }{}h_H{}^\beta {}_{\rho ]\sigma
}\!-S_{HV}^V{}^\sigma \!_{\beta [\alpha }{}h_V^T{}_{\rho ]}{}^\beta
\!_\sigma {}
\end{array}
\label{div.int.cond}
\end{equation}

\subsubsection{Projections of Contracted Curvatures}

Obtain the $HH$-projection of the Ricci curvature tensor from Eqs. (\ref
{R.hhhh.decomp},\ref{R.hvhv.decomp}) and take its complement to obtain a
result which has been the basis for singularity theorems in general
relativity:
\begin{equation}
\begin{array}{c}
R\left[ _{HH}\right] {}_{\alpha \beta }=R_{HH\,}^H{}_{\alpha \beta
}+D_{V\,\sigma }\,h_{H\,}{}^\sigma {}_{\alpha \beta }{}+D_{H\,\beta
}\,\theta _V^T{}_\alpha \\
+h_H{}^\rho {}_{\alpha \sigma }{}\left( h_H^T{}_\rho {}^\sigma \!_\beta
{}-\,S_{HV}^H{}^\sigma \!_{\beta \rho }{}\right) -\,h_{H\,}{}^\sigma
{}_{\alpha \beta }{}\theta _H^T{}_\sigma +h_V^T{}_\alpha {}^\rho \!_\sigma
\,{}S_{HV}^V{}^\sigma \!_{\beta \rho }{}
\end{array}
\label{Ricci.hh}
\end{equation}
\begin{equation}
\begin{array}{c}
R\left[ _{VV}\right] {}_{\alpha \beta }=R_{VV\,}^V{}_{\alpha \beta
}+D_{H\,\sigma }\,h_{V\,}{}^\sigma {}_{\alpha \beta }{}+D_{V\,\beta
}\,\theta _H^T{}_\alpha \\
+h_V{}^\rho {}_{\alpha \sigma }{}\left( h_V^T{}_\rho {}^\sigma \!_\beta
{}-\,S_{VH}^V{}^\sigma \!_{\beta \rho }{}\right) -\,h_{V\,}{}^\sigma
{}_{\alpha \beta }{}\theta _V^T{}_\sigma +h_H^T{}_\alpha {}^\rho \!_\sigma
\,S_{VH}^H{}^\sigma \!_{\beta \rho }{}
\end{array}
\label{Ricci.vv}
\end{equation}
Here $R_{VV\,\rho \alpha }^V$ is the intrinsic Ricci curvature which is
associated with the projection tensor field $V$.

By contracting this last result with the projected metric tensor $%
g^{VV\alpha \beta }$ one finds the {\em generalized Raychaudhuri equation}:
\begin{equation}
\begin{array}{c}
g^{VV}{}^{\alpha \beta }{}R\left[ _{VV}\right] {}_{\alpha \beta
}{}=R_{VV\,}^{V~}{}+D_{V\alpha }{}\theta _H^T{}^\alpha {}+D_{H\beta
}{}\theta _V{}^\beta {}-\theta _V^T{}_\sigma {}\theta _V{}^\sigma
{}+h_V^T{}_\sigma {}^{\beta \rho }{}h_{V\,}{}^\sigma {}_{\rho \beta }{} \\
+Q_V^{VV}{}^{\alpha \rho }{}_\alpha {}\theta _H^T{}_\rho
{}+Q_H^{VV}{}^{\alpha \rho }{}_\beta {}h_{V\,}{}^\beta {}_{\rho \alpha
}{}-h_V{}^\beta {}^\alpha {}_\sigma {}S_{VH}^V{}^\sigma \!_{\alpha \beta
}{}+h_H^T{}^\alpha {}^\beta {}_\sigma {}S_{VH}^H{}^\sigma \!_{\alpha \beta
}{}
\end{array}
\label{Raychaud.eqn}
\end{equation}
which governs the evolution of geodesic congruences.

The mixed projection can be obtained from Eqs. (\ref{R.hhhv.decomp},\ref
{R.hvvv.decomp})
\begin{equation}
\begin{array}{c}
R\left[ _{HV}\right] {}_{\alpha \beta }=R_{VH\,}^H{}_{\alpha \beta
}-D_{V\,\sigma }\,h_{V\,}^T{}_\alpha {}^\sigma \!_\beta {}+D_{V\,\beta
}\,\theta _V^T{}_\alpha \\
-h_H{}^\rho {}_{\alpha \sigma }{}\left( h_V\!^\sigma {}_{\rho \beta
}{}+\,S_{VV}^H{}^\sigma \!_{\beta \rho }{}\right) +\theta _H^T{}_\sigma
{}h_V^T{}_\alpha {}^\sigma \!_\beta {}+h_V^T{}_\alpha {}^\rho \!_\sigma
\,S_{VV}^V{}^\sigma \!_{\beta \rho }{}
\end{array}
\label{Ricci.hv}
\end{equation}
where $R_{VH\,\rho \alpha }^H=R_{VH\,\rho ~~\alpha \beta }^{H\quad ~~\beta }$
is the cross-projected Ricci tensor. The projections $R\left[ _{HH}\right] $
and $R\left[ _{VH}\right] $ may be obtained by taking the complements of
these results.

Forming the scalar curvature requires the use of a tensor $g^{\alpha \beta }$%
. This tensor plays the role of a metric on one-forms. However, it need not
be invertible or covariantly constant. In terms of the intrinsic and
cross-projected metric tensors, the scalar curvature is
\begin{equation}
\begin{array}{c}
R=R_{HH\,}^H{}+R_{VH\,}^H{}+R_{HV\,}^V{}+R_{VV\,}^V{} \\
+g^{HH}{}^{\alpha \beta }{}\left[ D_{H\,\beta }\,{}\theta _{V\,}^T{}_\alpha
{}+D_{V\,\sigma }{}\,h_{H\,}{}^\sigma {}_{\alpha \beta }{}-h_{H\,}{}^\rho
{}_{\alpha \sigma }{}\left( \,S_{HV}^H{}^\sigma \!_{\beta \rho
}{}-h_H^T{}_\rho {}^\sigma \!_\beta {}\right) \right. \\
\left. -h_{H\,}{}^\sigma {}_{\alpha \beta }{}\theta _{H\,}^T{}_\sigma
{}+h_V^T{}_\alpha {}^\sigma \!_\rho \,{}\,S_{HV}^V{}^\rho \!_{\beta \sigma
}{}\right] \\
+g^{HV}{}^{\alpha \beta }{}\left[ D_{V\beta }\,{}\theta _{V\,}^T{}_\alpha
{}-D_{V\sigma }\,{}h_V^T{}_\alpha {}^\sigma \!_\beta \,{}-h_{H\,}{}^\rho
{}_{\alpha \sigma }{}\left( S_{VV}^H{}^\sigma \!_{\beta \rho
}{}+h_{V\,}{}^\sigma {}_{\rho \beta }{}\right) \right. \\
\left. +\theta _{H\,}^T{}_\sigma {}h_V^T{}_\alpha {}^\sigma \!_\beta
{}+h_V^T{}_\alpha {}^\sigma \!_\rho {}\,S_{VV}^V{}^\rho \!_{\beta \sigma
}{}\right] \\
+g^{VH}{}^{\alpha \beta }{}\left[ D_{H\beta }\,{}\theta _{H\,}^T{}_\alpha
{}-D_{H\sigma }\,{}h_H^T{}_\alpha {}^\sigma \!_\beta \,{}-h_{V\,}{}^\rho
{}_{\alpha \sigma }{}\left( S_{HH}^V{}^\sigma \!_{\beta \rho
}{}+h_{H\,}{}^\sigma {}_{\rho \beta }{}\right) \right. \\
\left. +\theta _{V\,}^T{}_\sigma {}h_H^T{}_\alpha {}^\sigma \!_\beta
\,{}+h_H^T{}_\alpha {}^\sigma \!_\rho {}\,S_{HH}^H{}^\rho \!_{\beta \sigma
}{}\right] \\
+g^{VV}{}^{\alpha \beta }{}\left[ D_{V\,\beta }\,{}\theta _{H\,}^T{}_\alpha
{}+D_{H\,\sigma }\,{}h_{V\,}{}^\sigma {}_{\alpha \beta }{}-h_{V\,}{}^\rho
{}_{\alpha \sigma }{}\left( \,S_{VH}^V{}^\sigma \!_{\beta \rho
}{}-h_V^T{}_\rho {}^\sigma \!_\beta {}\,\right) \right. \\
\left. -h_{V\,}{}^\sigma {}_{\alpha \beta }{}\theta _{V\,}^T{}_\sigma
{}+h_H^T{}_\alpha {}^\sigma \!_\rho \,{}\,S_{VH}^H{}^\rho \!_{\beta \sigma
}{}\right]
\end{array}
\label{Scal.crv.prjct}
\end{equation}

\subsection{Projections of Curvature Identities}

\subsubsection{Unprojected Identities}

The curvature tensor obeys the usual identities. From the Jacobi identity
which ensures consistency of covariant derivatives acting on functions and
their gradients, one finds the {\em torsion Bianchi identity}:%
$$
R_{[\gamma }{}^\rho {}_{\mu \nu ]}{}+\nabla _{[\gamma }{}S^{\,\rho }{}_{\mu
\nu ]}{}+S^{\,\rho }{}_{\sigma [\gamma }{}S^{\,\sigma }{}_{\mu \nu ]}{}=0.
$$
Here, I am using the usual index bracket notation to indicate total
antisymmetrization. Because the curvature and torsion are already
antisymmetric in their last two indexes, the antisymmetrization just
generates three terms with the indexes cyclically permuted. The Jacobi
identity for covariant derivatives acting on vector or form fields yields
the {\em curvature Bianchi identity} in the form%
$$
\nabla _{[\alpha }{}R_{\mid \rho \mid }{}^\gamma {}_{\mu \nu ]}{}+R_\rho
{}^\gamma {}_{\sigma [\alpha }{}S^{\,\sigma }{}_{\mu \nu ]}=0.
$$
Again I am using brackets to indicate antisymmetrization with vertical bars
around indexes which are not included. The definition of metricity and the
definition of curvature provide still another identity%
$$
R^{(\gamma \rho )}{}_{\alpha \beta }{}=\nabla _{[\alpha }{}Q^{\gamma \rho
}{}_{\beta ]}{}+\frac 12Q^{\gamma \rho }{}_\sigma {}S^{\,\sigma }{}_{\alpha
\beta }{}.
$$
The parentheses indicate symmetrization.

\subsubsection{Projected Torsion Bianchi Identities}

The above identities could be analyzed by projecting each identity in all
possible ways and expressing the results in terms of the intrinsic and
cross-projection objects which have been introduced so far. Contemplate this
task briefly and notice that it will generate a very large number of terms,
many of which will eventually cancel. A much more efficient procedure is to
start over with the Jacobi identities for the intrinsic and extrinsic
projected derivatives of functions and vectors and proceed directly to find
the identities obeyed by the intrinsic and cross torsions and curvatures.
All of the resulting identities turn out to have a common structure so that
it is easiest to write a single general expression with variable projection
labels before discussing where the individual identities come from. For
projection tensors $X,Y,Z,W$ we will establish the identity:
\begin{equation}
\begin{array}{c}
R_{X\,Y}^{Z\!W}{}_\gamma {}^\rho {}_{\mu \nu }{}+R_{Z\,X}^{Y\!W}{}_\nu
{}^\rho {}_{\gamma \mu }{}+R_{Y\,Z}^{X\!W}{}_\mu {}^\rho {}_{\nu \gamma
}{}+D_{Z\gamma }\,{}S_{XY}^W{}^\rho {}_{\mu \nu }{}+D_{Y\nu
}\,{}S_{ZX}^W{}^\rho {}_{\gamma \mu }{}+D_{X\mu }\,S_{YZ}^W{}^\rho {}_{\nu
\gamma }{} \\
-S_{ZH}^W{}^\rho {}_{\gamma \sigma }{}\,S_{XY}^H{}^\sigma {}_{\mu \nu
}{}-S_{YH}^W{}^\rho {}_{\nu \sigma }{}\,S_{ZX}^H{}^\sigma {}_{\gamma \mu
}{}-S_{XH}^W{}^\rho {}_{\mu \sigma }{}\,S_{YZ}^H{}^\sigma {}_{\nu \gamma }{}
\\
-S_{ZV}^W{}^\rho {}_{\gamma \sigma }{}\,S_{XY}^V{}^\sigma {}_{\mu \nu
}{}-S_{YV}^W{}^\rho {}_{\nu \sigma }{}\,S_{ZX}^V{}^\sigma {}_{\gamma \mu
}{}-S_{XV}^W{}^\rho {}_{\mu \sigma }{}\,S_{YZ}^V{}^\sigma {}_{\nu \gamma
}{}=0
\end{array}
\label{tor.bianchi}
\end{equation}

The Jacobi identity for intrinsic projected covariant derivatives acting on
a function $\phi $ establishes the consistency of the torsion definition and
yields two identities. From the coefficient of $D_{H\rho }{}\phi $ comes the
{\em intrinsic projected torsion Bianchi identity} corresponding to $\left(
X,Y,Z,W\right) =\left( H,H,H,H\right) $ in the above general expression.
{}From the coefficient of $D_{V\rho }\phi $ comes another identity
corresponding to $\left( X,Y,Z,W\right) =\left( H,H,H,V\right) $. This
expression is worth writing out and commenting on.
\begin{equation}
\begin{array}{c}
D_{H\gamma }\,{}S_{HH}^V{}^\rho {}_{\mu \nu }{}+D_{H\nu }\,{}S_{HH}^V{}^\rho
{}_{\gamma \mu }{}+D_{H\mu }\,S_{HH}^V{}^\rho {}_{\nu \gamma }{} \\
-S_{HH}^V{}^\rho {}_{\gamma \sigma }{}\,S_{HH}^H{}^\sigma {}_{\mu \nu
}{}-S_{HH}^V{}^\rho {}_{\nu \sigma }{}\,S_{HH}^H{}^\sigma {}_{\gamma \mu
}{}-S_{HH}^V{}^\rho {}_{\mu \sigma }{}\,S_{HH}^H{}^\sigma {}_{\nu \gamma }{}
\\
-S_{HV}^V{}^\rho {}_{\gamma \sigma }{}\,S_{HH}^V{}^\sigma {}_{\mu \nu
}{}-S_{HV}^V{}^\rho {}_{\nu \sigma }{}\,S_{HH}^V{}^\sigma {}_{\gamma \mu
}{}-S_{HV}^V{}^\rho {}_{\mu \sigma }{}\,S_{HH}^V{}^\sigma {}_{\nu \gamma
}{}=0
\end{array}
\label{twist.bianchi}
\end{equation}
When one specializes the general expression to this case, all of the
intrinsic and cross-projected curvature terms are missing because this
choice of projection labels gives $XW=YW=ZW=0$. Notice that the expression
is linear and homogeneous in the generalized twist tensor $S_{HH}^V{}^\rho
{}_{\mu \nu }{}$ and can be viewed as a consequence of that object's
definition in terms of the antisymmetric derivative of $H$. The complements
of these identities yield projected torsion Bianchi identities corresponding
to $\left( X,Y,Z,W\right) =\left( V,V,V,V\right) ,\left( V,V,V,H\right) $.

Similarly, the Jacobi identity for mixed intrinsic and extrinsic derivatives
acting on a function yields the {\em cross-projected torsion Bianchi
identities} which correspond to the cases $\left( X,Y,Z,W\right) =\left(
H,H,V,H\right) ,\left( H,H,V,V\right) $ the remaining identities,
corresponding to $\left( X,Y,Z,W\right) =\left( V,V,H,V\right) ,\left(
V,V,H,H\right) $ can be found by taking the complements of these. Since the
identities are symmetric under cyclic permutations of the labels $X,Y,Z$,
the proof of the general expression is complete.

\subsubsection{Projected Curvature Bianchi Identities}

The simplest way to decompose the curvature Bianchi identities is to start
with the Jacobi identity for projected covariant derivatives acting on
vectors or forms and use the definitions of the intrinsic and cross
curvature tensors to evaluate all of the commutators. From the Jacobi
identity for the intrinsic projected derivatives $D_{H\nu }$ acting on a
form $\eta $ with $\eta (P)$$\in H^{*}T_P$, one finds three identities, two
of which are torsion Bianchi identities which were found above and one is
new --- the {\em intrinsic projected curvature Bianchi identity}:%
$$
\begin{array}{c}
D_{H\gamma }{}R_{HH}^H{}_\delta {}^\rho {}_{\mu \nu }{}+S_{HH}^H{}^\sigma
{}_{\mu \nu }{}R_{HH}^H{}_\delta {}^\rho {}_{\sigma \gamma
}{}+S_{HH}^V{}^\sigma {}_{\mu \nu }{}R_{VH}^H{}_\delta {}^\rho {}_{\sigma
\gamma }{} \\
+D_{H\nu }{}R_{HH}^H{}_\delta {}^\rho {}_{\gamma \mu }{}+S_{HH}^H{}^\sigma
{}_{\gamma \mu }{}R_{HH}^H{}_\delta {}^\rho {}_{\sigma \nu
}{}+S_{HH}^V{}^\sigma {}_{\gamma \mu }{}R_{VH}^H{}_\delta {}^\rho {}_{\sigma
\nu }{} \\
+D_{H\mu }{}R_{HH}^H{}_\delta {}^\rho {}_{\nu \gamma }{}+S_{HH}^H{}^\sigma
{}_{\nu \gamma }{}R_{HH}^H{}_\delta {}^\rho {}_{\sigma \mu
}{}+S_{HH}^V{}^\sigma {}_{\nu \gamma }{}R_{VH}^H{}_\delta {}^\rho {}_{\sigma
\mu }{}=0
\end{array}
$$
The complement of this expression yields the corresponding identity for the
projection tensor $V$.

The Jacobi identities for mixed intrinsic and extrinsic projected covariant
derivatives each yield one new projected Bianchi identity. All of these
identities follow the same pattern as the intrinsic identity above. In terms
of the variable projection labels $X,Y,Z,W$ the identities are:%
$$
\begin{array}{c}
D_{X\gamma }{}R_{ZY}^W{}_\delta {}^\rho {}_{\mu \nu }{}+S_{ZY}^H{}^\sigma
{}_{\mu \nu }{}R_{HX}^W{}_\delta {}^\rho {}_{\sigma \gamma
}{}+S_{ZY}^V{}^\sigma {}_{\mu \nu }{}R_{VX}^W{}_\delta {}^\rho {}_{\sigma
\gamma }{} \\
+D_{Y\nu }{}R_{XZ}^W{}_\delta {}^\rho {}_{\gamma \mu }{}+S_{XZ}^H{}^\sigma
{}_{\gamma \mu }{}R_{HY}^W{}_\delta {}^\rho {}_{\sigma \nu
}{}+S_{XZ}^V{}^\sigma {}_{\gamma \mu }{}R_{VY}^W{}_\delta {}^\rho {}_{\sigma
\nu }{} \\
+D_{Z\mu }{}R_{YX}^W{}_\delta {}^\rho {}_{\nu \gamma }{}+S_{YX}^H{}^\sigma
{}_{\nu \gamma }{}R_{HZ}^W{}_\delta {}^\rho {}_{\sigma \mu
}{}+S_{YX}^V{}^\sigma {}_{\nu \gamma }{}R_{VZ}^W{}_\delta {}^\rho {}_{\sigma
\mu }{}=0
\end{array}
$$
{}From the identity for two intrinsic derivatives and one extrinsic derivative
acting on a form-field which assigns forms in $H^{*}T_P$, one finds the {\em %
cross-projected curvature Bianchi identity} corresponding to the projection
labels $\left( X,Y,Z,W\right) =\left( H,H,V,H\right) $. The Jacobi identity
for two extrinsic derivatives and one intrinsic derivative acting on a
form-field yields the cross-projected curvature Bianchi identity
corresponding to $\left( X,Y,Z,W\right) =\left( H,V,V,H\right) $ and so on.
There are six such cross-projected curvature Bianchi identities.

\subsubsection{Projected Curvature-Metricity Identities}

Begin with the intrinsic curvature definition for vectors $v\in HT_P$
$$
v^\rho R_{HH}^H{}_\rho {}^\gamma {}_{\alpha \beta }{}=\left( \left[
D_{H\beta }{},D_{H\alpha }{}\right] -S_{HH}^H{}^\rho {}_{\alpha \beta
}{}D_{H\rho }{}-S_{HH}^V{}^\rho {}_{\alpha \beta }{}D_{V\rho }{}\right)
v^\gamma {}
$$
and take $v^\rho {}=g^{\rho \delta }{}\xi _\delta {}=g^{HH}{}^{\rho \delta
}{}\xi _{H\delta }{}+g^{HV}{}^{\rho \delta }{}\xi _{V\delta }{}$. Use the
product rule for the derivatives to obtain an expression in which the
curvature operator acts directly on the forms $\xi _{H\delta }{}$ and $\xi
_{V\delta }{}$. Equating the coefficients of $\xi _{H\delta }{}$ yields the
identity

\begin{equation}
\begin{array}{c}
R_{HH}^H{}_\rho {}^\gamma {}_{\alpha \beta }{}g^{HH}{}^{\rho \delta
}{}+R_{HH}^H{}_\rho {}^\delta {}_{\alpha \beta }{}g^{HH}{}^{\gamma \rho }{}
\\
=D_{H\alpha }{}Q_H^{HH}{}^{\gamma \delta }{}_\beta {}-D_{H\beta
}{}Q_H^{HH}{}^{\gamma \delta }{}_\alpha {}+S_{HH}^H{}^\rho {}_{\alpha \beta
}{}Q_H^{HH}{}^{\gamma \delta }{}_\rho {}+S_{HH}^V{}^\rho {}_{\alpha \beta
}{}Q_V^{HH}{}^{\gamma \delta }{}_\rho {}
\end{array}
\label{cmethhhh}
\end{equation}
While the coefficients of $\eta _{V\delta }$ yield
\begin{equation}
\begin{array}{c}
R_{HH}^H{}_\rho {}^\gamma {}_{\alpha \beta }{}g^{HV}{}^{\rho \delta
}{}+R_{HH}^V{}_\rho {}^\delta {}_{\alpha \beta }{}g^{HV}{}^{\gamma \rho }{}
\\
=D_{H\alpha }{}Q_H^{HV}{}^{\gamma \delta }{}_\beta {}-D_{H\beta
}{}Q_H^{HV}{}^{\gamma \delta }{}_\alpha {}+S_{HH}^H{}^\rho {}_{\alpha \beta
}{}Q_H^{HV}{}^{\gamma \delta }{}_\rho {}+S_{HH}^V{}^\rho {}_{\alpha \beta
}{}Q_V^{HV}{}^{\gamma \delta }{}_\rho {}.
\end{array}
\label{cmethhhv}
\end{equation}

By beginning with the cross-projected curvature definitions, one finds still
more identities of this sort. There are sixteen such identities in all. All
of the definitions and operations which go into deriving these identities
have exactly the same structure, differing only in the projection labels. In
terms of variable projection labels, $X,Y,Z,W$ all of the identities may be
obtained from the expression:
\begin{equation}
\begin{array}{c}
R_{XY}^Z{}_\rho {}^\gamma {}_{\alpha \beta }{}g^{ZW}{}^{\rho \delta
}{}+R_{XY}^W{}_\rho {}^\delta {}_{\alpha \beta }{}g^{ZW}{}^{\gamma \rho }{}
\\
=D_{Y\alpha }{}Q_X^{ZW}{}^{\gamma \delta }{}_\beta {}-D_{X\beta
}{}Q_Y^{ZW}{}^{\gamma \delta }{}_\alpha {}+S_{XY}^H{}^\rho {}_{\alpha \beta
}{}Q_H^{ZW}{}^{\gamma \delta }{}_\rho {}+S_{XY}^V{}^\rho {}_{\alpha \beta
}{}Q_V^{ZW}{}^{\gamma \delta }{}_\rho {}
\end{array}
\label{cmeth.gen}
\end{equation}

The structural similarity of these curvature-metricity identities disappears
when assumptions are made about the metric, the metricity, or the torsion.
For example, normal projection tensors are characterized by $%
g^{VH\,}=0,\quad Q_H^{VH\,}=0,\quad Q_V^{VH\,}=0$ so that the identities
which correspond to $\left( Z,W\right) =\left( V,H\right) $ become empty.

\section{Familiar Applications}
\label{applications}

\subsection{Perfect Fluid Thermodynamics}

An earlier paper\cite{prjctn1} discussed projection tensor fluid dynamics,
so I will not go into much detail here. However, the earlier paper made very
restrictive assumptions about the geometry --- a normal projection tensor in
a torsion and metricity-free spacetime. It is interesting to note that those
assumptions are unnecessary and do not even simplify the discussion.

A perfect fluid has the stress-energy tensor%
$$
T^{\,\mu }{}_\nu {}=p_HH^{\,\mu }{}_\nu {}+p_VV^{\,\mu }{}_\nu {}
$$
where $p_H=p$ is the pressure and $p_V=-\rho $ where $\rho $ is the
mass-energy density. The projection tensor $H$ and its complement $V$ are
normal projection tensors as described in section \ref{settings} or this
paper. However, this fact is not needed. A projection decomposition of the
conservation law $\nabla _\mu {}T^{\,\mu }{}_\nu {}=0$ follows directly from
the contraction of Eq. (\ref{prjctn.grad.decomp}) which, with the
projection identities obeyed by the projection curvatures, yields
$$
\nabla _\mu {}H^{\,\mu }{}_\nu {}=\theta _H^T{}_\nu {}-\theta _V^T{}_\nu {}
$$
and its complement so that%
$$
\nabla _\mu {}T^{\,\mu }{}_\nu {}=D_{H\,\nu }\,{}p_H+p_H\left( \theta
_H^T{}_\nu {}-\theta _V^T{}_\nu {}\right) +D_{V\,\nu }\,{}p_V+p_V\left(
\theta _V^T{}_\nu {}-\theta _H^T{}_\nu {}\right)
$$
The conservation law then implies the equation%
$$
D_{H\,\nu }\,{}p_H+\left( p_V-p_H\right) {}\theta _V^T{}_\nu {}=0
$$
and its complement. As was discussed in the earlier paper, these equations
are indeed Euler's equation and the equation of continuity for fluid flow.

A curious result of the earlier paper is that the complementary pairing of
pressure and energy density could be extended to a pairing of all the
thermodynamic potentials by insisting that the ''Tds'' equations of
thermodynamics should be invariant with respect to the complement operation.
The resulting pairs are $T_H=$ temperature, $T_V=-$ baryon density, $s_H=$
entropy density, $s_V=$ chemical potential. In terms of these definitions,
the chemical potential is defined by the relation%
$$
p_H-p_V=T_Hs_H-T_Vs_V
$$
and the first law of thermodynamics is%
$$
dp_V=s_VdT_V-T_Hds_H.
$$
The law of baryon conservation has the same form as the continuity equation,
but without a pressure term:%
$$
D_{V\,\nu }\,T_V-T_V\,\theta _H^T{}_\nu {}=0.
$$
Complementation symmetry takes on the appearance of magic at this point
because the complement of this last equation is recognizable as another
valid thermodynamic relation, the general relativistic thermal equilibrium
condition --- the red-shifted temperature is a constant. As the earlier
paper discussed, other thermodynamic relations may also be derived in
complementary pairs.

I am a bit surprised that the simple projection-geometry form of the fluid
thermodynamic equations does not depend on the use of normal projection
tensors to construct the fluid stress-energy. In fact, it is also unaffected
by the properties of the connection --- metricity and torsion make no
difference at all. Evidently, the geometrical definition of the divergence $%
\theta _{H\;\nu }^T$ of a projection tensor field corresponds closely to
what the physics of fluids requires and other aspects of the geometry do not
play direct roles.

\subsection{Normal Projection Tensor Fields and Torsionless
Metric-compatible Connections}

\subsubsection{The Simplifications}

The curvature decomposition and the various torsion and curvature identities
are strongly affected by the normality of the projection tensor field as
well as by the properties of the connection. For a torsion-free connection,
the intrinsic torsion is zero but the cross-projected torsions are not:
\begin{equation}
S_{HH}^V{}^\rho {}_{\mu \nu }{}=2\omega _H{}^\rho {}_{\mu
\nu }{},\qquad S_{HV}^H{}^\rho {}_{\mu \nu }{}=h_H^T{}_\nu {}^\rho {}_\mu {}
\label{c.tor.spec}
\end{equation}
For a normal projection tensor field, $h=h^T$, and $g^{VH}=g^{HV}=0$. With
these specializations, the divergence integrability condition and the
generalized Raychaudhuri equation become:
$$
0=D_{H[\alpha }{}\theta _{V\,\rho ]}{}+D_{V\beta }{}\omega _H{}^\beta
{}_{\rho \alpha }{}-h_H{}_\beta {}^\sigma \!_{[\alpha }{}h_H{}^\beta
{}_{\rho ]\sigma }{}-h_V{}_{[\rho }{}^\beta \!_{\left| \sigma \right|
}{}h_V{}_{\alpha ]}{}^\sigma \!_\beta {}
$$
$$
g^{VV}{}^{\alpha \rho }{}R\left[ _{VV}\right] {}_{\alpha \rho
}{}=R_{VV\,}^{V~}+D_{V\alpha }{}\theta _H{}^\alpha +D_{H\beta }{}\theta
{}_V{}^\beta -\theta _{V\,\sigma }{}\theta _V{}^\sigma -h_H{}_\alpha
{}^\sigma \!_\beta {}h_H{}^\alpha {}^\beta \!_\sigma {}
$$
The projections of the Riemann and Ricci tensors as well as the various
projected Bianchi identities also simplify.

\subsubsection{Timelike Geodesic Congruences: Dust Clouds}

A cloud of freely falling particles is represented by a congruence of
timelike geodesics --- the world-lines of the particles. I will have nothing
new to say about this well-understood system. However, its very familiarity
makes it a useful illustration of projection tensor geometry.

Begin with the projection tensor field $V^\alpha \!_\beta {}=-u^\alpha
{}u_\beta $ where $u^\alpha $ is the four-velocity of the particles. The
accelerations of the particles are described by the curvature $h_V$ of this
projection tensor field. For particles in free fall,
$$
h_V^T{}^\alpha {}_{\gamma \delta }{}=h_V{}^\alpha {}_{\gamma \delta
}{}=V_{\gamma \delta }{}a^\alpha {}=0.
$$
The vanishing of the entire projection curvature tensor is a consequence of
the one-dimensional nature of the projection subspaces $VT_P$. The
projection curvature associated with $H$ does not vanish and is related to
the usual twist, shear, and divergence as follows:
\begin{equation}
h_H^T{}^\gamma {}_{\alpha \beta }{}=h_H{}^\gamma {}_{\alpha
\beta }{}=k_{\alpha \beta }u^\gamma .
\label{dust.crv}
\end{equation}
\begin{equation}
k_{\alpha \beta }=\sigma _{\alpha \beta }{}+\omega _{\alpha
\beta }{}+\frac 13\theta H_{\alpha \beta }{}
\label{shear.etc}
\end{equation}

In this situation, the divergence integrability condition, Eq. (\ref
{div.int.cond}), becomes just%
$$
D_{V\beta }\omega _H{}^\beta {}_{\rho \alpha }=h_H{}_\beta {}^\sigma
\!_{[\alpha }{}h_H{}^\beta {}_{\rho ]\sigma }{}
$$
Substitute $h_H=\theta _H+\omega _H$ on the right-side of this equation and
obtain a result which is manifestly linear in the vorticity:
\begin{equation}
D_{V\beta }\omega _H{}^\beta {}_{\rho \alpha }{}=\theta
_H{}_\beta {}^\sigma \!_\alpha {}\omega _H{}^\beta {}_{\rho \sigma
}{}+\omega _H{}_\beta {}^\sigma \!_\alpha {}\theta _H{}^\beta {}_{\rho
\sigma }{}
\label{vort.evolve1}
\end{equation}
The projected derivative $D_{V\beta }$ generates just the usual Fermi
derivative along the particle world-lines so this result is just the usual
evolution equation for the vorticity. In detail, the definition of the
projected derivative as well as Eqs. (\ref{dust.crv},\ref{shear.etc})
give the result%
$$
H^\sigma \!_\rho {}H^\tau \!_\alpha V^\beta \!_\mu V^\delta \!_\beta
{}\nabla _\delta \left( \omega _{\sigma \tau }{}u^\mu \right) =\frac
12\left( k^\sigma {}_\alpha {}u_\beta k_{\rho \sigma }{}u^\beta -k^\sigma
{}_\rho {}u_\beta k_{\alpha \sigma }{}u^\beta \right)
$$
which collapses to
$$
H^\sigma \!_\rho {}H^\tau \!_\alpha u^\mu {}\nabla _\mu {}\omega _{\sigma
\tau }=\frac 12\left( k_{\alpha \sigma }k^\sigma {}_\rho -k_{\rho \sigma
}k^\sigma {}_\alpha \right)
$$
or, recognizing the Fermi derivative on the left and using $k=\theta +\omega
$ on the right,
\begin{equation}
\dot \omega \left[ _{HH}\right] {}_{\alpha \beta
}=\theta _{\alpha \sigma }{}\omega ^\sigma {}_\rho {}+\omega _{\alpha \sigma
}{}\theta ^\sigma {}_\rho
\label{vort.evolve2}
\end{equation}

One remarkable (and well-known) thing about this evolution equation for the
vorticity is that it depends only on the local anisotropic expansion rate of
the cloud. There is no direct dependence on the spacetime geometry. Looking
back at the projection of the Riemann tensor which gave rise to this result,
Eq. (\ref{R.hvhv.decomp}), it can be seen that the antisymmetrization
in Eq. (\ref{div.int.cond}) eliminates the curvature term when it has
the usual index symmetries. In the presence of torsion, the Riemann tensor
would not have all of the usual index symmetries and there would be a direct
contribution of the spacetime geometry to the evolution of the vorticity.

Another remarkable thing about this evolution equation is that its essential
structure can be read from the general projection tensor form in equation (%
\ref{vort.evolve1}). Specializing the equation still further to obtain
Eq. (\ref{vort.evolve2}) does not yield any new insights.

Evolution equations for the shear and divergence can be obtained from the
projections of the Ricci tensor and Einstein's field equations in their
trace-reversed and projected form:%
$$
R_{\mu \nu }{}=T_{\mu \nu }{}-\frac 12Tg_{\mu \nu }
$$
For example, the generalized Raychaudhuri equation needs the projection%
$$
R\left[ _{VV}\right] {}_{\alpha \beta }{}=\left( T_{\mu \nu }{}-\frac
12Tg_{\mu \nu }\right) u^\mu {}u^\nu {}u_\alpha {}u_\beta =-\frac 12\left(
\rho +\sum p_i\right) V_{\alpha \beta }
$$
where $\rho $ is the total mass-energy density of all the matter present and
$\sum p_i$ is the sum of the principal pressures. With all of the
specializations which apply to a cloud of free particles, the generalized
Raychaudhuri equation reduces to%
$$
-\frac 12\left( \rho +\sum p_i\right) =D_{V\alpha }\theta _H{}^\alpha
-h_H{}_\alpha {}^\sigma \!_\beta {}h_H{}^\alpha {}^\beta \!_\sigma {}.
$$
Notice that the one-dimensional nature of $V$ has been used to eliminate the
term $R_{VV\,}^{V~}$. A short calculation gives this equation in the usual
form%
$$
\dot \theta =-\frac 12\left( \rho +\sum p_i\right) -\sigma ^2-\frac 13\theta
^2+\omega ^2
$$
which shows that $\dot \theta <0$ whenever the cloud has no vorticity and
the strong energy condition is satisfied --- The cloud tends to collapse
because gravitation is attractive.

\subsubsection{String Clouds}

In an earlier paper\cite{prjctn1}, I noted that a projection-tensor
formulation of fluid dynamics can be applied directly to string fluids by a
simple change in the dimensionality of the projection tensors which are
used. Here I apply the same technique to freely falling strings. Instead of
a one-dimensional projection onto the world-lines of particles, let $V$ be a
two-dimensional projection onto the timelike world-sheets of freely falling
strings. Consider a cloud of such strings in which the world-sheets do not
cross each other and repeat the analysis of the previous section to
determine how such a cloud can evolve.

First, think about ordinary particles again. When $V$ projects onto the
world-lines of freely falling particles, the corresponding projection
curvature is zero. This result can be obtained by noting that the divergence
form $\theta _V{}_\alpha {}$ can be expressed as the rate of change of the
line element along a world-line as one moves from one dust-particle to
another. The particles move so as to extremize the lengths of their world
lines, which leads directly to the condition
\begin{equation}
\theta _V{}_\alpha =0.
\label{div.zero}
\end{equation}
Writing this condition in terms of the usual variables, it becomes just $%
a_\alpha =0$ --- the particles are unaccelerated. Thus, the vanishing of the
projection divergence captures the essential equation of motion of a cloud
of freely falling particles. The additional consequence that the entire
projection curvature $h_V{}^\mu {}_\nu {}_\alpha $ vanishes is an accidental
consequence of the low dimensionality of the projection tensor $V$.

Now turn to freely falling strings. In this case, $V$ projects onto a
timelike two-dimensional surface. Keep Eq. (\ref{div.zero}) as the
essential dynamical condition and see if it makes sense. Here, the
divergence $\theta _V{}_\alpha $ expresses the rate of change of the
timelike area element from one string to the next. Thus, Eq. (\ref
{div.zero}) corresponds to strings which move so as to have extremal area
world-sheets --- just what is usually assumed for Goto-Nambu bosonic strings
and certainly a reasonable generalization of timelike geodesics.\cite
{string.area.action,string.area.action.a,string.area.action.b} Because the
projection curvature tensor $h_V{}^\mu {}_\nu {}_\alpha $ now has more
components than its divergence, it will not necessarily vanish. However,
there is an additional requirement because the strings are assumed to be
extended objects which hold together and sweep out surfaces. From the
projection-tensor version of Frobenius's Theorem, this requirement means
that
\begin{equation}
\omega _V{}^\alpha {}_{\mu \nu }{}=0.
\label{vort.zero}
\end{equation}
Combining the two requirements (Eqs. (\ref{div.zero},\ref{vort.zero}))
yields the most general form which the $V$ projection curvature of a freely
falling string-cloud can have:
\begin{equation}
h_V{}^\alpha {}_{\mu \nu }{}=\sigma _V{}^\alpha {}_{\mu
\nu }{}.
\label{vcrv.shear}
\end{equation}

The divergence integrability condition, Eq. (\ref{div.int.cond}), for
this string cloud becomes%
$$
D_{V\beta }\omega _H{}^\beta {}_{\rho \alpha }{}-h_H{}_\beta {}^\sigma
\!_{[\alpha }{}h_H{}^\beta {}_{\rho ]\sigma }{}-h_V{}_{[\rho }{}^\beta
\!_{\left| \sigma \right| }{}h_V{}_{\alpha ]}{}^\sigma \!_\beta {}=0
$$
The last term in this expression vanished for particle clouds because the $V$%
-projection curvature vanished in that case. Here, the term is again zero
because of Eq. (\ref{vort.zero}). As a result, we simply get equation (%
\ref{vort.evolve1}) again. The interpretation of the equation is slightly
different because the first index on the vorticity tensor $\omega _H{}^\alpha
{}_{\mu \nu }{}$ can now take two different values.

To interpret the projection curvature of a string-cloud, choose an adapted
orthonormal coordinate system with the spacelike basis vector $e_1=s$ and a
timelike vector $e_0=u$ tangent to the string world-sheets. The projection
curvature $H$ is then found to have components%
$$
h_H{}^0{}_{\gamma \delta }{}=k_{\gamma \delta },\qquad h_H{}^1{}_{\gamma
\delta }{}=-b_{\gamma \delta }
$$
where $k_{\gamma \delta }=\nabla u\left[ _{HH}\right] _{\gamma \delta }$ is
the familiar projected gradient of the timelike fluid flow vector field $u$
while $b_{\gamma \delta }=\nabla s\left[ _{HH}\right] _{\gamma \delta }$ is
the corresponding object --- the projected gradient of the spacelike string
tangent vector field $s$ --- for a $t=$const. snapshot of the string-cloud.
Thus, there is both a spacelike {\em curl} $b_{\left[ \mu \nu \right]
}=-\omega _H{}^1{}_{\mu \nu }{}$ and a timelike {\em vorticity} $k_{\left[
\mu \nu \right] }=\omega _{\mu \nu }=\omega _H{}^0{}_{\mu \nu }{}$ and these
two tensors do not have separate evolution equations. Inspecting equation (%
\ref{vort.evolve1}) reveals that it is an evolution equation for the string
vorticity and has a term proportional to the gradient of the string curl.
Physically, this makes perfect sense: The strings can change their vorticity
by winding and unwinding.

The generalized Raychaudhuri equation does not prove to be quite so useful
for strings as it is for particles. For a string cloud, it takes the form
$$
V^{\alpha \rho }R{}_{\alpha \rho }=R_{VV\,}^{V~}+D_{V\alpha }\theta
_H{}^\alpha -h_H{}_\alpha {}^\sigma \!_\beta {}h_H{}^\alpha {}^\beta
\!_\sigma {}.
$$
In the adapted orthonormal holonomic frame on string world-sheets, the
expression becomes%
$$
\begin{array}{c}
\dot \theta ^0=-p_2-p_3-R_{VV\,}^{V~}-n\cdot \nabla \theta ^1-\left[ \sigma
^0\right] ^2-\frac 12\left( \theta ^0\right) ^2+\left[ \omega ^0\right] ^2
\\
+\left[ \sigma ^1\right] ^2+\frac 12\left( \theta ^1\right) ^2-\left[ \omega
^1\right] ^2
\end{array}
$$
Unlike the particle case, there are no reasonable conditions under which the
timelike divergence $\theta ^0$ is guaranteed to be decreasing. The terms in
the expression do make physical sense, however. Gravity continues to be
purely attractive, tending to collapse the string cloud, but acts only
through the transverse principal pressures. Positive intrinsic curvature of
the strings tend to collapse the cloud. If the spatial divergence, $-\theta
^1$ of the strings decreases as one moves in the positive direction along
the strings, then the string tension tends to collapse the cloud just as
elementary Newtonian physics would suggest. As with a particle cloud,
timelike shear and divergence collapse the cloud while timelike vorticity
tends to expand it.

\subsubsection{Spacelike 3+1 Initial Value Analysis}

Take $H$ to be the normal projection onto a spacelike hypersurface $\Sigma $
and project Einstein's equations%
$$
G_{\mu \nu }=R_{\mu \nu }-\frac 12Rg_{\mu \nu }=T_{\mu \nu }
$$
in all possible ways. The curvature projection equations in this paper make
this a relatively straightforward process.

The only unfamiliar feature of the calculation is the appearance of
cross-projected curvature terms such as $R_{HV}^H$. Some of these terms are
easily disposed of by using the one-dimensional nature of $VT_P$. The
definition of the cross-projected curvatures, yield $R_{VV}^X=0$ and the
curvature-metricity relations in Eq. (\ref{cmeth.gen}) with the
metricity set to zero yield $R_{XY}^V=0$ in this one-dimensional case. to
dispose of the cross-projected curvature $R_{HV}^H$, turn to the projected
torsion Bianchi identity in Eq. (\ref{tor.bianchi}) with $\left(
X,Y,Z,W\right) =\left( V,H,H,H\right) $. With the unprojected torsion set to
zero, and some help from Eq. (\ref{c.tor.spec}), this identity becomes%
$$
\begin{array}{c}
R_{VH}^H{}_\gamma {}^\rho {}_{\mu \nu }{}+R_{HV}^H{}_\nu {}^\rho {}_{\gamma
\mu }=D_{H\gamma }{}h_H{}_\mu {}^\rho {}_\nu {}-D_{H\nu }{}h_H{}_\mu {}^\rho
{}_\gamma {} \\
+h_H{}_\sigma {}^\rho {}_\gamma {}h_V{}_\nu {}^\sigma {}_\mu {}-h_H{}_\sigma
{}^\rho {}_\nu {}h_V{}_\gamma {}^\sigma {}_\mu {}
\end{array}
$$
Contract this identity and notice that the curvature-metricity identity
requires $R_{HV}^H{}_\rho {}^\rho {}_{\gamma \mu }=0.$ The resulting
identity shows how to express the cross-Ricci-curvature in terms of
projection curvatures.%
$$
R_{VH}^H{}_{\gamma \mu }{}=D_{H\gamma }\theta _H{}_\mu {}-D_{H\rho
}{}h_H{}_\mu {}^\rho {}_\gamma {}+h_H{}_\sigma {}^\rho {}_\gamma
{}h_V{}_\rho {}^\sigma {}_\mu {}-\theta _H{}_\sigma {}h_V{}_\gamma {}^\sigma
{}_\mu {}
$$

Once the cross-curvature terms have been eliminated from the projections of
the Ricci curvature tensor, the rest of the task is familiar. It is
important to realize, however, that the relation
$$
h_V^T{}^\alpha {}_{\gamma \delta }{}=h_V{}^\alpha {}_{\gamma \delta
}{}=V_{\gamma \delta }{}\theta _V{}^\alpha
$$
is needed to produce the usual simple results. This relation as well as the
simplifications already used to express the cross-projected curvature terms
depend on the one-dimensional nature of the projected tangent space $VT_P$ .
Because these relations are obviously not symmetrical between $H$ and $V$,
the resulting expressions will not have complementation symmetry.

The scalar curvature expression in Eq. (\ref{Scal.crv.prjct})
simplifies to just

$$
\begin{array}{c}
R=R_{HH\,}^H+2D_{V\,\sigma }\,\theta _H{}^\sigma +2D_{H\,\sigma }\,\theta
_V{}^\sigma -2\,\theta _V{}^\sigma \theta _{V\sigma } \\
-\,\theta _H{}^\sigma \theta _{H\sigma }-h_H{}^\beta {}^\sigma {}_\rho
{}h_H{}_\beta {}^\rho {}_\sigma {}
\end{array}
$$
and the Einstein tensor projections follow from the Ricci tensor projections
given in Eqs. (\ref{Ricci.hh},\ref{Ricci.vv},\ref{Ricci.hv}). The
results are the familiar ones in an only slightly unfamiliar form:
$$
G\left[ _{HV}\right] {}_{\alpha \beta }=-D_{H\rho }{}p{}_\beta {}^\rho
{}_\alpha {}
$$
$$
g^{\alpha \beta }G\left[ _{VV}\right] {}_{\alpha \beta }{}=-\frac 12\left(
R_{HH\,}^H{}+p{}^\beta {}^\rho \!_\sigma {}\,p{}_\beta {}^\sigma \!_\rho
{}-\frac 14\,p{}^\sigma p_\sigma \right)
$$
\begin{equation}
\begin{array}{c}
G\left[ _{HH}\right] {}_{\alpha \beta }{}=D_{V\,\sigma }\,p{}^\sigma
{}_{\alpha \beta }{}+U_{\alpha \beta }{}-UH_{\alpha \beta }{}+\frac
12p_\sigma {}p_{\,}{}^\sigma {}_{\alpha \beta }{} \\
{}+\frac 12\left( p{}^\tau {}^\sigma {}_\rho {}p{}_\tau {}^\rho {}_\sigma
{}-\frac 12p{}_\tau {}p^\tau {}{}\right) H_{\alpha \beta
}+G_{HH\,}^H{}_{\alpha \beta }
\end{array}
\label{prjctd.dynam.eqs}
\end{equation}
where I define%
$$
p{}_\gamma {}^\alpha {}_\beta {}=h_H{}_\gamma {}^\alpha {}_\beta {}-H^\alpha
{}_\beta {}\theta _H{}_\gamma {}
$$
and%
$$
U_{\alpha \beta }{}=D_{H\,\beta }\,\theta _V{}_\alpha -\theta _V{}_\alpha
\,\theta _V{}_\beta {}.
$$
It is interesting to note that the divergence integrability condition,
equation(\ref{div.int.cond}), ensures that the tensor $U_{\alpha \beta }{}$
is symmetric and that $\theta _V{}_\alpha $ is the gradient of a scalar
potential.

As has been discussed in many places, using many different approaches\cite
{iv-projectns}, it is evident that four of Einstein's equations contain no
timelike derivatives of $p_{\alpha \beta }{}^\gamma $ and serve only to
constrain the initial value data while the remaining six can be regarded as
providing the time derivatives which are needed to evolve the field. I will
not complete the projection-tensor geometry version of the discussion here.
Instead, I will just note what is left to be done at this point: (1) Express
the projection curvatures in terms of Lie derivatives of the intrinsic
metric along a timelike curve congruence. (2) Make explicit the dependence
on the arbitrary choice of curve congruence (The ADM approach uses the
''lapse function'' and the ''shift vector'' for this purpose.\cite
{lapse.shift.gr.iv}$^{,}$\cite{lapse.shift.gr.iv.a} Here we have the freedom
to let $V$ project directly onto the curve congruence by relaxing the
restriction to normal projections. In that case, $g^{VV}{}$ plays the role
of the lapse and $g^{HV}{}$ contains the shift vector). (3) Organize the
resulting equations of motion into one or another constrained Hamiltonian
form. (4) Discuss conditions which can be imposed in order to constrain the
choice of timelike curve congruence.

\section{Discussion}

Although projection tensor techniques have often been used in general
relativity, they have always been restricted in peculiar ways: Useful in the
spacelike initial value problem of general relativity but not in the
characteristic initial value problem; Easily applied to the spacelike
initial value Einstein equations but notoriously difficult to apply to the
remaining, dynamical Einstein equations\cite{gr.iv.dynam.eqs}; Useful for
spacelike projections in fluids and initial value problems but not for
timelike projections; Useful when the co-dimension is one but less useful
otherwise. This paper shows that the restrictions have been the result of
several ''missing puzzle pieces'' which are needed to perform
straightforward calculations in projection tensor geometry. These pieces are:

\begin{itemize}
\item  The transpose projection curvature tensor.

\item  The cross-projected torsion and curvature tensors.
\end{itemize}

To see what can happen, compare the calculation of Eq. (\ref
{prjctd.dynam.eqs}) with the calculations which appear in the literature\cite
{gr.iv.dynam.eqs}. The calculation here is made easy by using the
cross-projected torsion Bianchi identities to eliminate the cross-projected
curvature tensor terms. Even though the cross-projected torsion and
curvature do not occur in the final answer, they play an essential role in
getting there and it is easy to see why the result can be tedious to obtain
without them.

In addition to filling in missing pieces, this paper has extended the
projection tensor approach to new situations and given it increased
flexibility in familiar situations. Thus, null hypersurfaces and the
characteristic initial value problem can now be fitted into the same
geometrical framework that has been used for spacelike hypersurfaces. The
intrinsic geometry classification of fluid-containing spacetimes developed
by Collins and Szafron\cite{intrinsic.geom.I,intrinsic.geom.II}$^{,}$ can
now be extended to higher dimensional cases and perhaps developed further in
other ways. Even the familiar 3+1 calculations can now be done in new ways
without resorting to the explicit use of coordinates. I expect to exploit
some of these opportunities in later papers.


\begin{thebibliography}{99}
\bibitem{curv.note}  My conventions agree with standard texts such as refs%
\cite{co-moving,iv-projectns} when the torsion and metricity vanish. For
non-zero torsion my notation is driven by the convention that indexes
associated with derivatives are always listed last. The most common
variation in the literature is to swap the first two indexes of the
curvature with the last two.
\bibitem{intrinsic.note}  The term ''intrinsic geometry'' is being used in a
more restricted sense here than Collins and Szafron use it in refs. \cite
{intrinsic.geom.I,intrinsic.geom.II,intrinsic.geom.III} .
\end{thebibliography}
\end{document}